\documentclass[aps,prb,twocolumn,showpacs,superscriptaddress]{revtex4-1}
\usepackage{graphicx}
\usepackage{blindtext}
\usepackage{amsmath}
\usepackage{amssymb}
\usepackage{colordvi}
\usepackage{mathrsfs}
\usepackage{bm}
\usepackage{bbold}
\usepackage{verbatim}
\usepackage{dcolumn}
\usepackage{bm}
\usepackage{epsfig}
\usepackage{physics}
\usepackage{makecell}
\usepackage[export]{adjustbox}
\usepackage[colorlinks,allcolors=blue]{hyperref}

\begin{document}      
	\title{Topological phase transition from periodic edge states in moiré superlattices}
	\author{Haonan Wang}
	\affiliation{Department of Physics, Washington University in St. Louis, St. Louis, Missouri 63130, USA}
	\author{Li Yang}%
	\affiliation{Department of Physics, Washington University in St. Louis, St. Louis, Missouri 63130, USA}
	\email[Correspondence author:~]{lyang@wustl.edu}
	\date{\today}     
        \begin{abstract}
            Topological mosaic pattern (TMP) can be formed in two-dimensional (2D) moiré superlattices, a set of periodic and spatially separated domains with distinct topologies give rise to periodic edge states on the domain walls. In this study, we demonstrate that these periodic edge states play a crucial role in determining global topological properties. By developing a continuum model for periodic edge states with $C_{6z}$ and $C_{3z}$ rotational symmetry, we predict that a global topological phase transition at the charge neutrality point (CNP) can be driven by the size of domain walls and moiré periodicity. The Wannier representation analysis reveals that these periodic edge states are fundamentally chiral $p_x\pm ip_y$ orbitals. The interplay between on-site chiral orbital rotation and neighboring hopping among chiral orbitals leads to band inversion and a topological phase transition. Our work establishes a general model for tuning local and global topological phases, paving the way for future research on strongly correlated topological flat minibands within topological mosaic patterns.
        \end{abstract}   
\maketitle

\section{Introduction} Moiré superlattices formed in twisted or lattice-mismatched van der Waals (vdW) materials present a highly tunable platform for engineering electronic correlation and topology. One celebrated example is the Mott insulator as well as unconventional superconductivity discovered in mgaic-angle twisted bilayer graphene (TBG) \cite{ref72,ref71}. Moreover, topological phases have been widely explored in twisted graphene systems \cite{ref4,ref5,ref6,ref7,ref37,ref38,ref39,ref40,ref41,ref42,ref43}, homo- and hetero-bilayers of twisted 2D transition metal dichalcogenides (TMDs) \cite{ref8,ref9,ref10,ref11,ref44,ref45,ref46,ref47}, and so on \cite {ref12,ref13,ref14,ref48}.
For small twist angles, local electronic states are slowly modulated over long-period moiré superlattices, and tuning stacking-dependent local properties may significantly alter global properties of the moiré pattern.

\begin{figure}[tb]
	\centering
    \begin{tabular}{l l}
      \includegraphics[width=0.28\textwidth,valign=t]{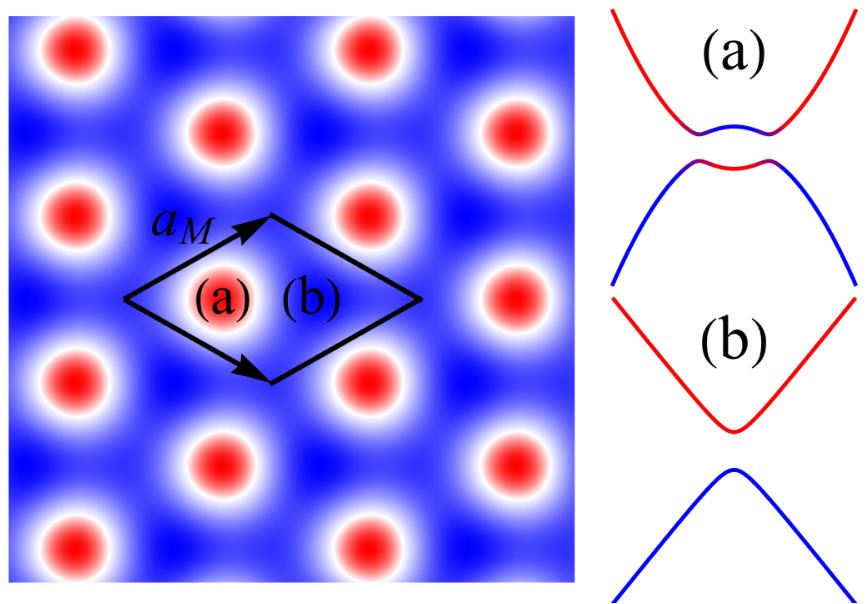} & 
      \includegraphics[width=0.185\textwidth,valign=t]{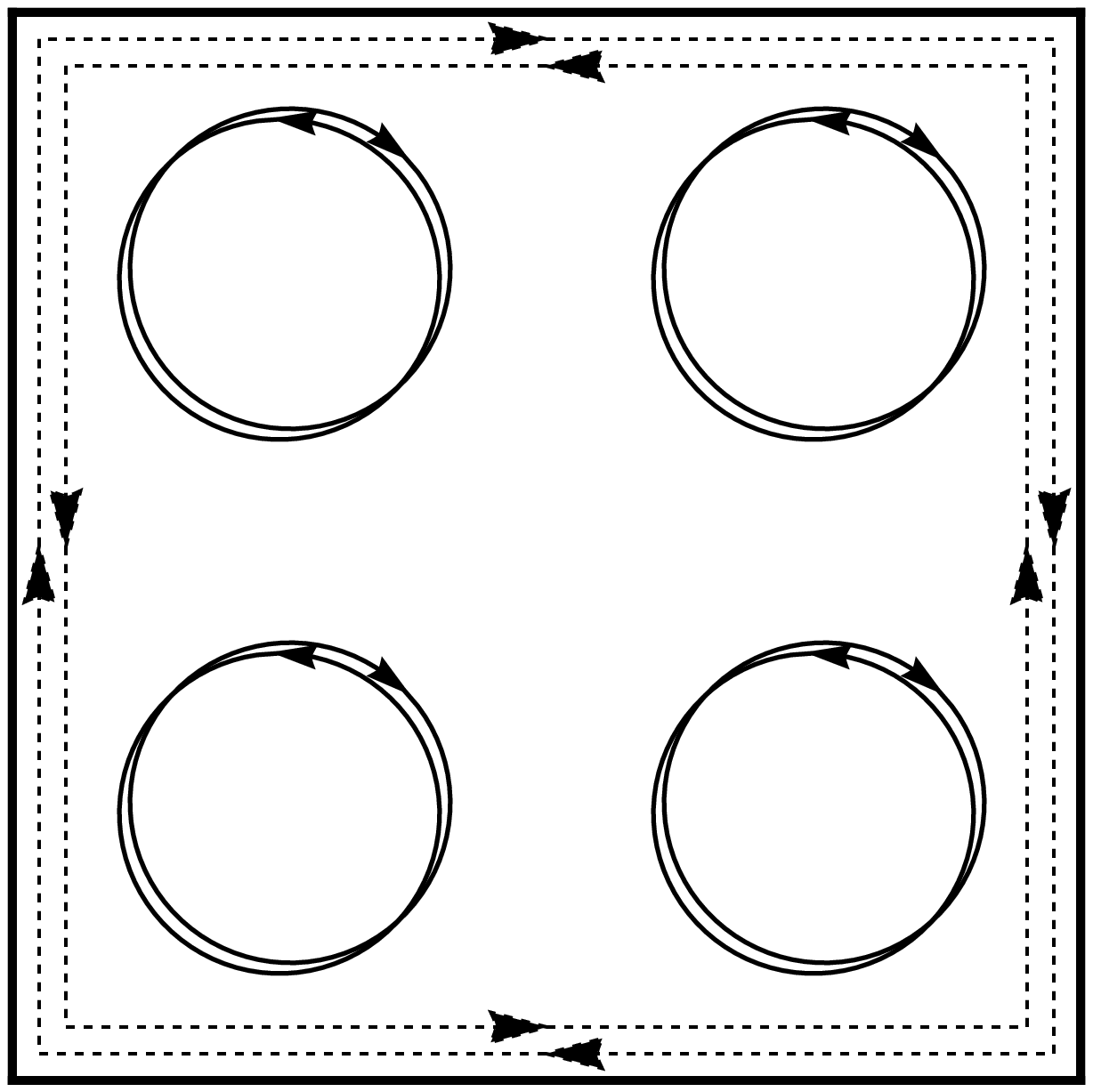} 
    \end{tabular}
	\caption{Left panel: topological mosaic pattern (TMP), including (a) TI domains with band inversion, and (b) NI domains without band inversion. The white circles stand for periodic edge states on TI/NI domain walls. Right panel: sketch of topological mosaic pattern. The black circles denote helical periodic edge states. The dashed lines indicate possible helical edge states at the boundary of the TMP.}
	\label{fig1}
\end{figure}

The formation of a topological mosaic pattern in long-period moiré superlattices is an intriguing phenomenon, as depicted in Fig. \ref{fig1}. The presence of different domains with distinct topologies in the 2D structure results in the emergence of periodic edge states along the domain walls, in accordance with the bulk-boundary correspondence \cite{ref2,ref22}. One way to create such a TMP is by twisting van der Waals (vdW) materials with stacking-dependent topology \cite{ref13,ref15,ref16,ref17,ref29,ref36,ref76}. For instance, in heterobilayer TMDs, the type-II band alignment for certain atomic configurations can be adjusted to fall into the inverted regime with a few hundred meV interlayer bias \cite{ref29,ref36}. TBG with a large interlayer bias also exhibit insulating AB and BA stacking domains with opposite valley Chern numbers \cite{ref54,ref55,ref70,ref73}. Furthermore, thin films of topological insulators (TI) such as bilayer Bi$_2$Te$_3$ or even magnetic TIs like MnBi$_2$Te$_4$ family can exhibit stacking-dependent topology without interlayer bias \cite{ref15,ref16,ref17,ref76}. Remarkably, their twisted counterparts can host topological flat moiré minibands and moiré-scale edge states, implying global TI phases for the TMP via doping \cite{ref12,ref13}. However, it is still unclear whether the topological flat minibands in these materials are connected to the emergent periodic edge states and how global TI phases can be controlled by adjusting the local TI phases at CNP, as depicted in Fig. \ref{fig1}. To address these questions, a model based on the periodic edge states is necessary.


In this paper, we present a continuum model for periodic edge states on domain walls in the topological mosaic pattern. Our findings reveal a topological phase transition at the charge neutrality point that can be driven by the size of domain walls and the moiré periodicity. Using maximally localized Wannier functions for the low-energy minibands, we establish that the topological phase transition arises from overlaps of the periodic edge states described by neighboring hopping between $p_x\pm ip_y$ orbitals on the triangular lattice. Additionally, we provide topological phase diagrams for $C_{6z}$ and $C_{3z}$-symmetric domain walls. To demonstrate the versatility of our model, we calculate moiré minibands of twisted bilayer Bi$_2$Te$_3$ (TBBT) and the domain wall network of TBG, both featuring $C_{3z}$-symmetric domain walls. Our model's findings align with results from some previous studies.

The remainder of this paper is organized as follows. The continuum model Hamiltonian for the periodic edge states is constructed, and the symmetry analysis for the model is also presented in Sec. II. The study of tunable band topology and topological phase diagram for $C_{6z}$,$C_{3z}$-symmetric domain walls are presented, and a tight-binding Hamiltonian based on the maximally localized Wannier orbitals for the low energy bands for $C_{6z}$-symmetric domain walls is constructed in Sec. III. Further discussion about the model is presented in Sec. IV. Some auxiliary material is
relegated to the Appendixes.

\section{Continuum Model Hamiltonian} As seen in Fig. \ref{fig1}, the TMP allows 1D edge states to reside on the domain walls. Recall that 1D edge states at the boundary of 2D TIs are closely related to Jackiw-Rebbi zero-mode, which appears as the zero energy soliton of Dirac equation \cite {ref2,ref18,ref19}. In other words, the low energy bound states can be solved from a massive Dirac equation to reside at the interface between two regions with opposite masses. Inspired by this idea, we modify the 2D Dirac equation to describe periodic edge states by considering the mass term as a real space function with moiré periodicity. Then domains with opposite masses represent topologically distinct areas in TMP. The modified Dirac Hamiltonian can be expressed as 
\begin{equation}\label{eq1}    
\begin{aligned}     
\mathcal{H}&=-i\hbar v\Big(\alpha_x\partial_x+\alpha_y\partial_y\Big)+\beta M(\mathbf r),
\end{aligned}
\end{equation}    
where $\alpha_x=\mathbf{1}_{s}\otimes\sigma_x$, $\alpha_y=s_z\otimes\sigma_y,\beta=\mathbf{1}_{s}\otimes\sigma_z$, satisfying Clifford algebra. Here $\boldsymbol{s}$ and $\boldsymbol{\sigma}$ are Pauli matrices that correspond to spin and orbital pseudospin degrees of freedom. Hence $M(\mathbf r)$ denotes an orbital Zeeman splitting. Here we note that i) the Hamiltonian in Eq. \eqref{eq1} is mathematically identical to the Bernevig-Hughes-Zhang (BHZ) model with a moiré scale oscillation in the mass term \cite{ref13,ref33}, ii) $M(\mathbf r)$ is time-reversal (TR) invariant so that the two Dirac cones in Eq. \eqref{eq1} could be gapped without breaking the TR symmetry \cite{ref1}, and iii) a similar model Hamiltonian has been used to describe surface states of TIs gapped by magnetic Zeeman moiré potential \cite{ref11,ref14,ref56}, which however breaks the TR symmetry. Up to the second-order Fourier approximation, $M(\mathbf r)$ with three-fold rotational symmetry about the $z$-axis ($C_{3z}$) can be expanded as
\begin{equation}   \label{eq2} 
\begin{aligned}     
    M(\mathbf r)=-m_0-2m_1\sum_{j=1,3,5}\cos(\mathbf G_j\cdot \mathbf r+\phi),
	\end{aligned}
\end{equation}
where the summation runs over the three nearest neighbor reciprocal lattices. $\mathbf G_j=4\pi/(\sqrt3a_M)(\cos{2\pi\frac{(j-1)}{3}},\sin{2\pi\frac{(j-1)}{3}})$ $(j=1,3,5)$. $m_0,m_1,\phi$ alter the landscape of the moiré potential $M(\mathbf r)$ while $a_M$ controls moiré periodicity. Note that $\mathcal{H}$ is TR invariant as $T\mathcal{H}_{\mathbf k}(\mathbf r)T^{-1}=\mathcal{H}_{-\mathbf k}(\mathbf r)$, $T=i\sigma_yK\otimes\mathbf{1}_{\sigma}$ with $T^2=-1$, where $K$ is the complex conjugation operator. It also preserves the particle-hole symmetry $S\mathcal{H}_{\mathbf k}(\mathbf r)S^{-1}=-\mathcal{H}_{\mathbf k}(\mathbf r)$ where $S=\mathbf{1}_{s}\otimes i\sigma_yK$, which implies the energy spectrum to be symmetric with respect to zero energy. Furthermore, $s_z$ is a good quantum number so $\mathcal{H}$ has two decoupled parts 
\begin{equation} \label{eq3}
\begin{aligned}  
     H^\uparrow_{\mathbf k}(\mathbf r)&=v(k_x\sigma_x+k_y\sigma_y)+M(\mathbf r)\sigma_z,\\
     H^\downarrow_{\mathbf k}(\mathbf r)&=H^{\uparrow *}_{\mathbf k}(\mathbf r),
     \end{aligned}
\end{equation}
where we have set $\hbar=1$. Hence the local spin Chern number for each part can be defined as
\begin{equation} \label{eq4}
\begin{aligned}  
        C_{local}=\int\frac{d\mathbf k}{4\pi}\frac{\mathbf d}{d^3}\cdot\bigg[\frac{\partial\mathbf d}{\partial k_x}\times\frac{\partial\mathbf d}{\partial k_y}\bigg]=\frac{M}{2\abs{M}}
     \end{aligned}
\end{equation}
where the vector $\mathbf d$ is defined by the Pauli matrix expansion of Eq. \eqref{eq3}, $H^\uparrow=(\mathbf d\cdot\boldsymbol\sigma)$. The local spin Chern number difference across the domain walls is 1, namely one helical
electronic channel residing on the domain walls. 

 Up to a scale $\mathcal{H}$ has three independent dimensionless parameters $m_0/m_1,\phi$ and $a_Mm_1/v$. For simplicity we set $m_1=v=1$ in our calculations. Bloch's theorem allows $\mathcal{H}$ to act on the spinor $\Psi_{n\mathbf k}(\mathbf r)=(\psi_{n\mathbf k}^{\uparrow}(\mathbf r),\psi_{n\mathbf k}^{\downarrow}(\mathbf r))^T$, where $\psi_{n\mathbf k}^{\uparrow}(\mathbf r)$ is a two-component Bloch's function, and the band structures can be computed by using the plane-wave expansion with truncated 169 plane wave basis that are sufficient for energy convergence. 

\begin{figure}[tb]
  \centering
  \setlength{\tabcolsep}{0pt}
    \begin{tabular}{l l}
      \includegraphics[width=0.25\textwidth,valign=t]{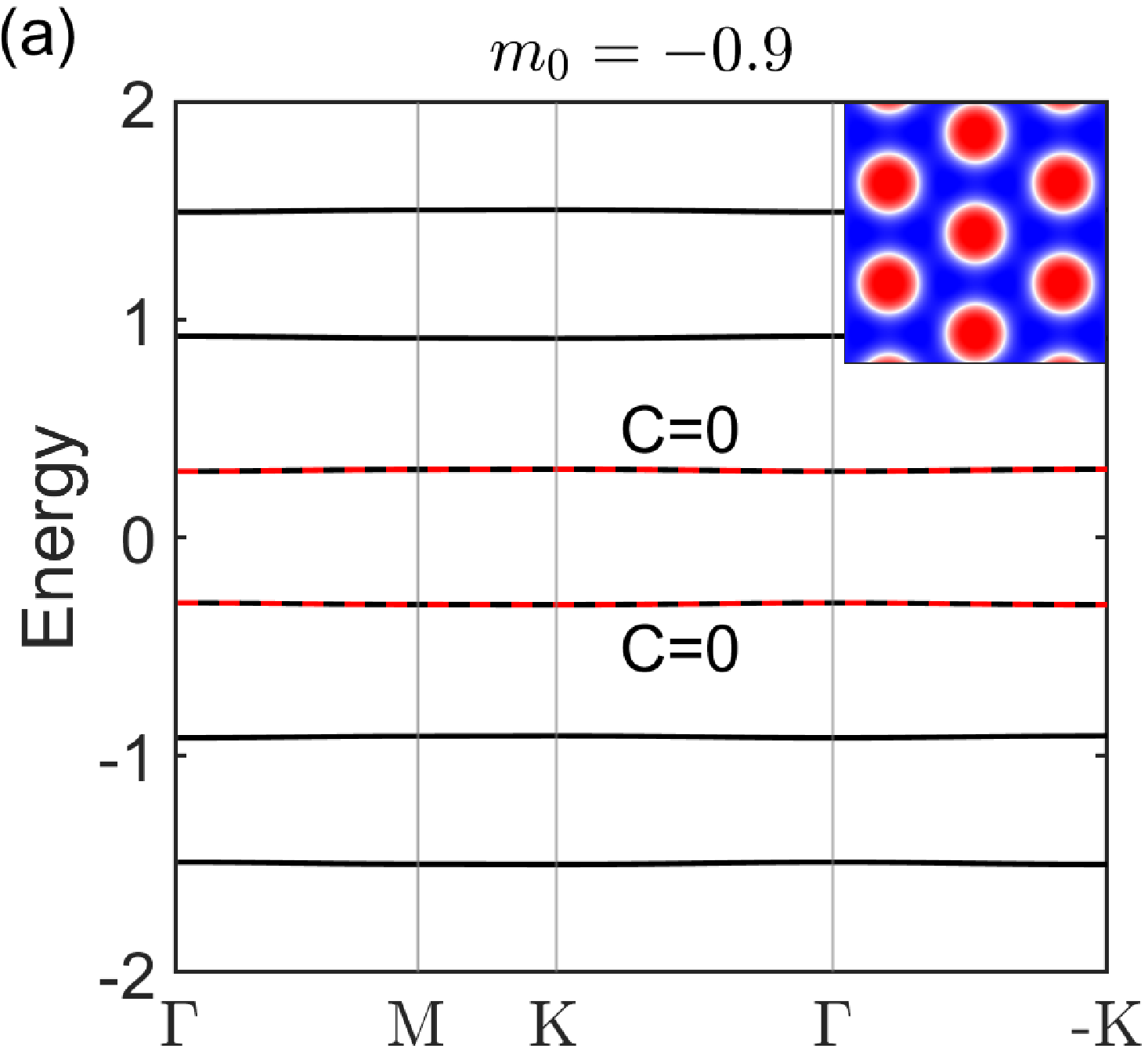} & \includegraphics[width=0.25\textwidth,valign=t]{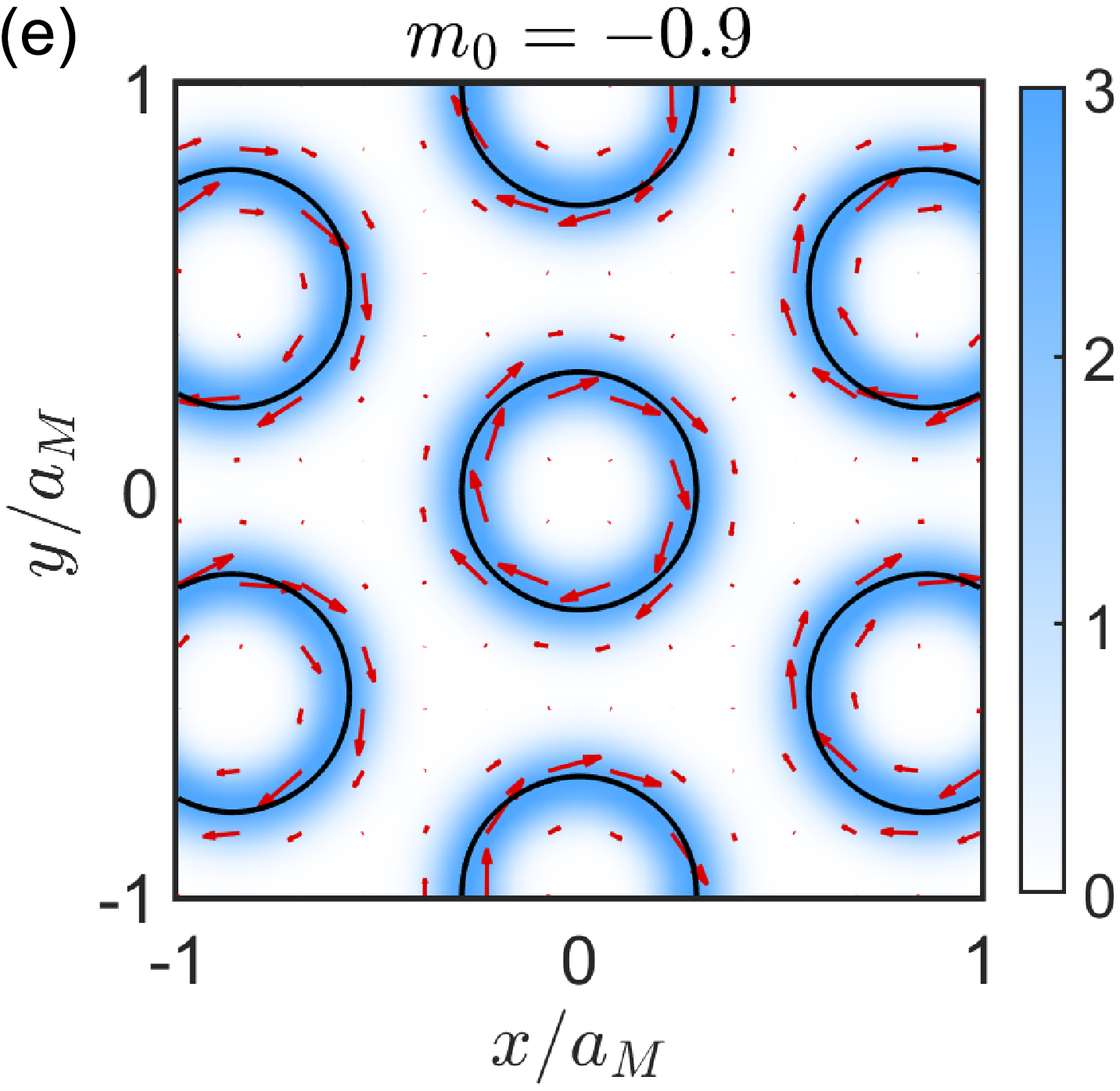} 
      \\[1.5\normalbaselineskip]
      \includegraphics[width=0.25\textwidth,valign=t]{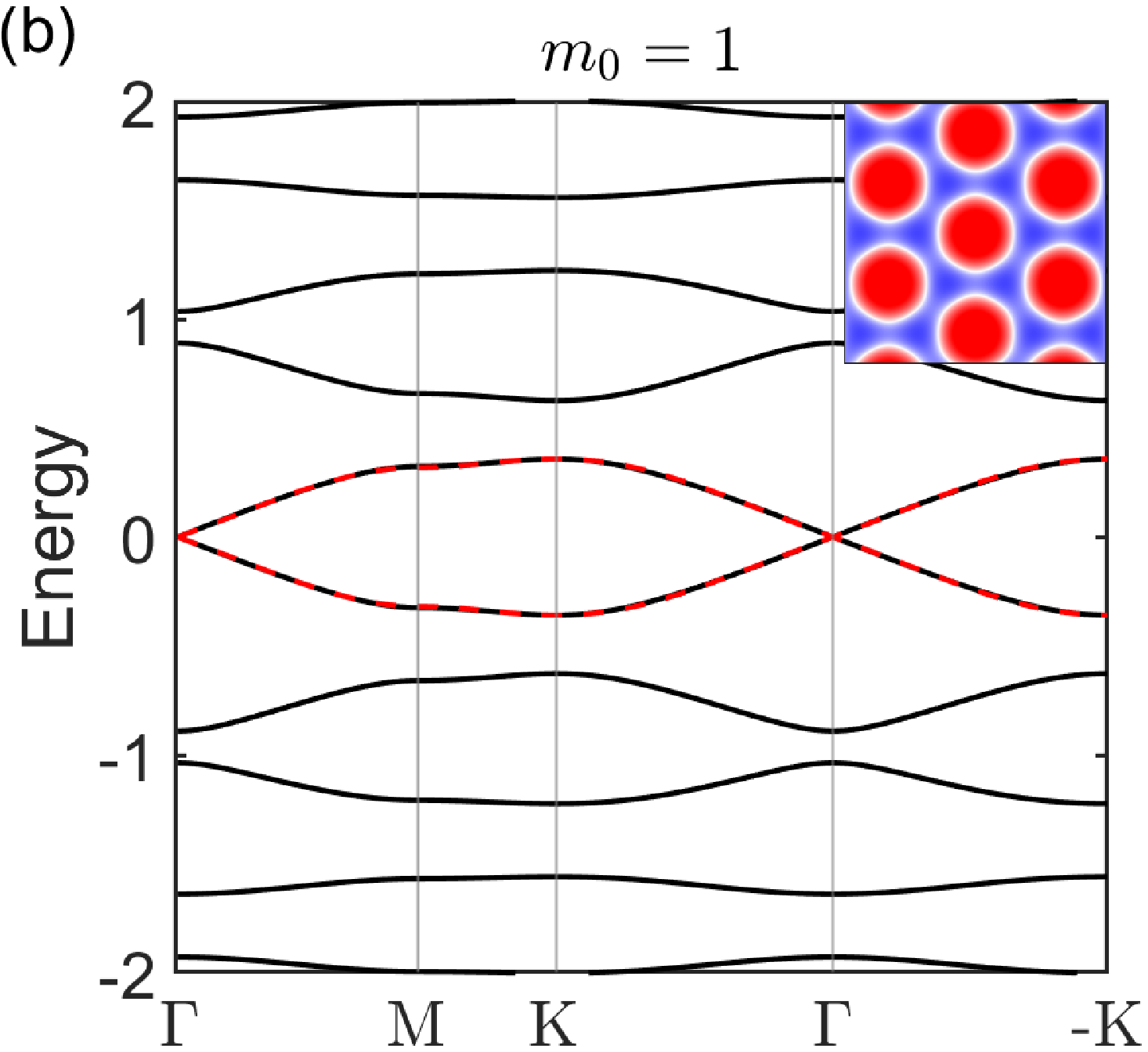} &
      \includegraphics[width=0.25\textwidth,valign=t]{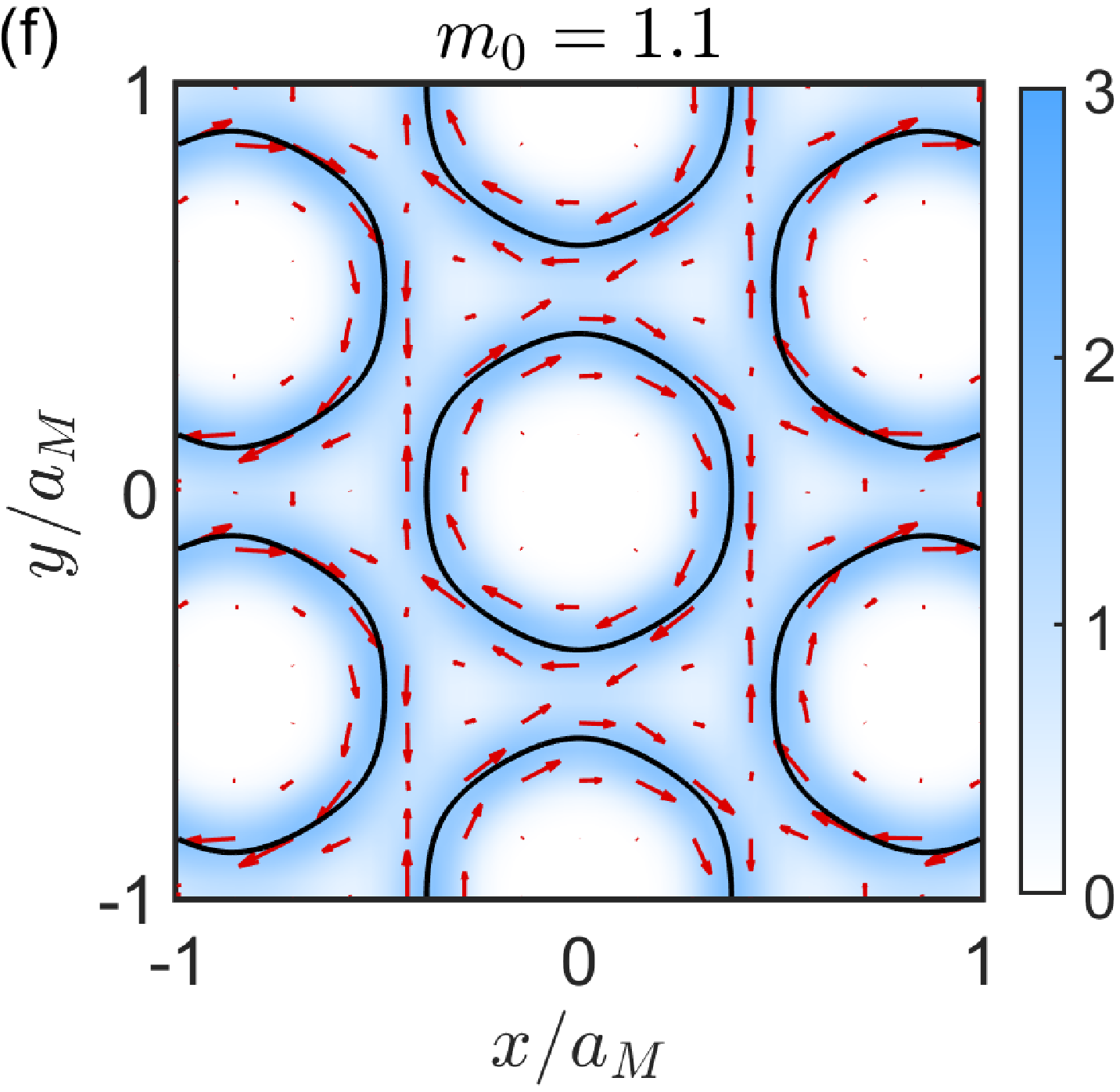} 
      \\[1.5\normalbaselineskip]
      \includegraphics[width=0.25\textwidth,valign=t]{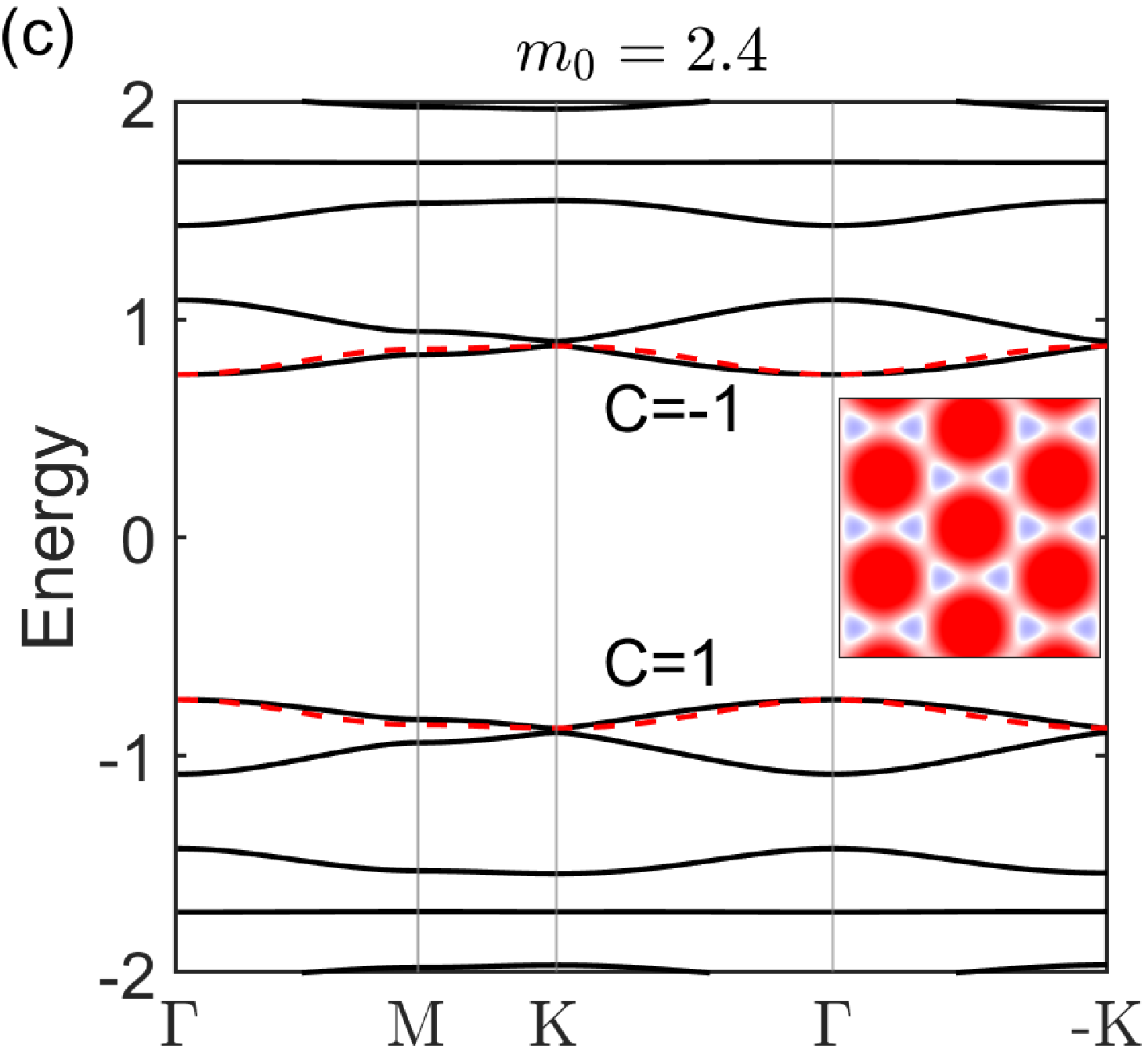} &
      \includegraphics[width=0.25\textwidth,valign=t]{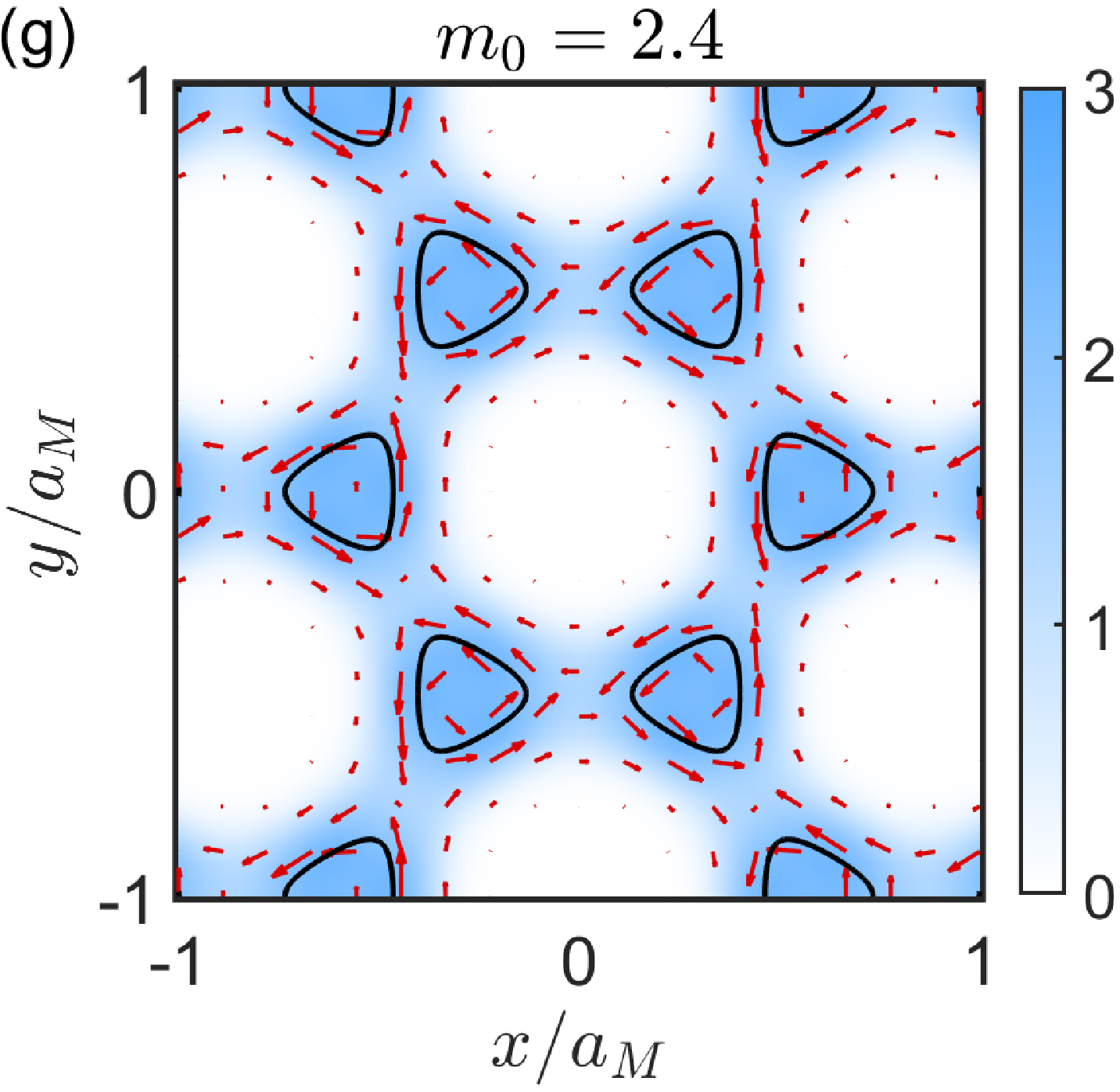}
      \\[1.5\normalbaselineskip]
      \includegraphics[width=0.25\textwidth,valign=t]{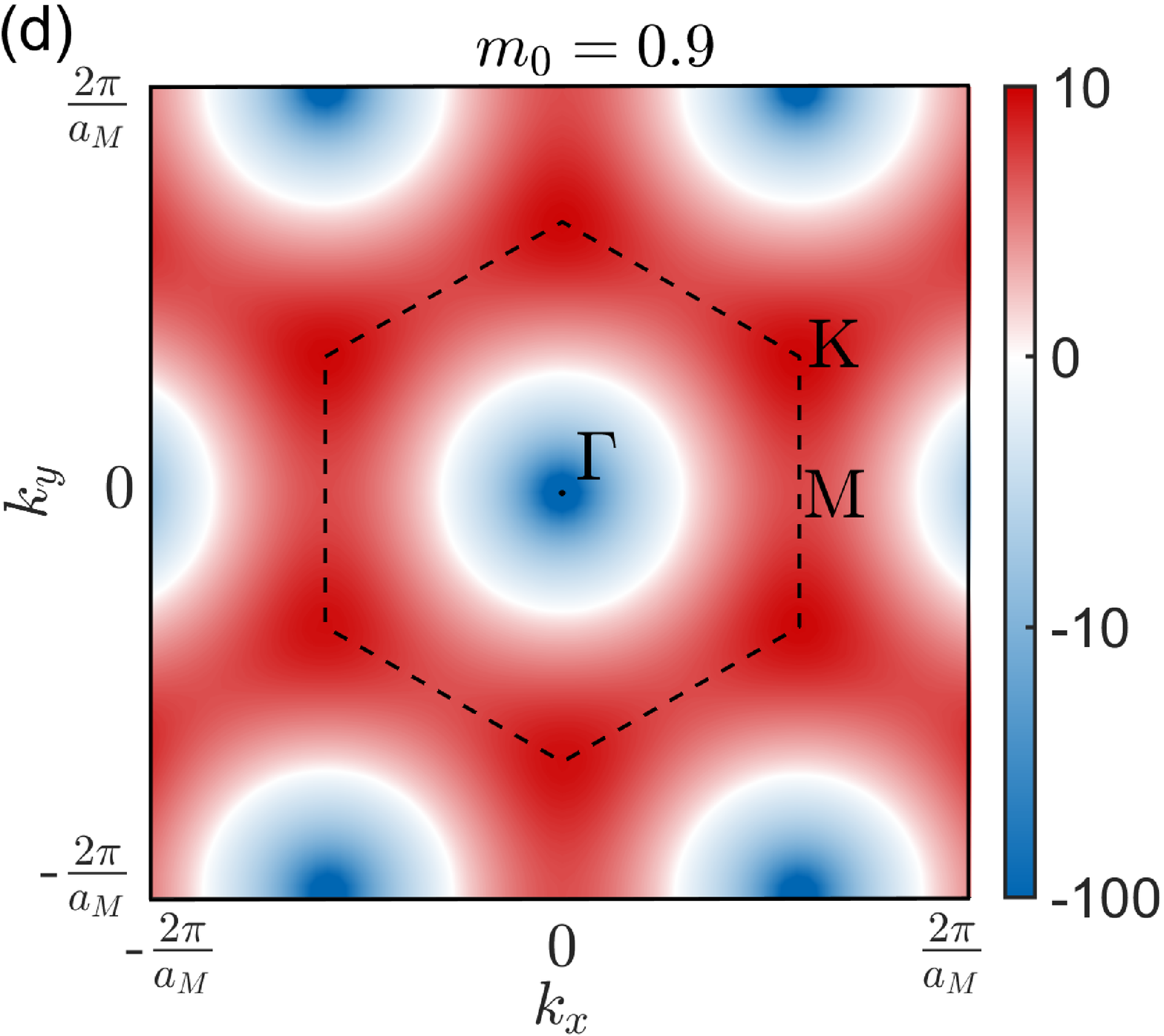} &
      \includegraphics[width=0.25\textwidth,valign=t]{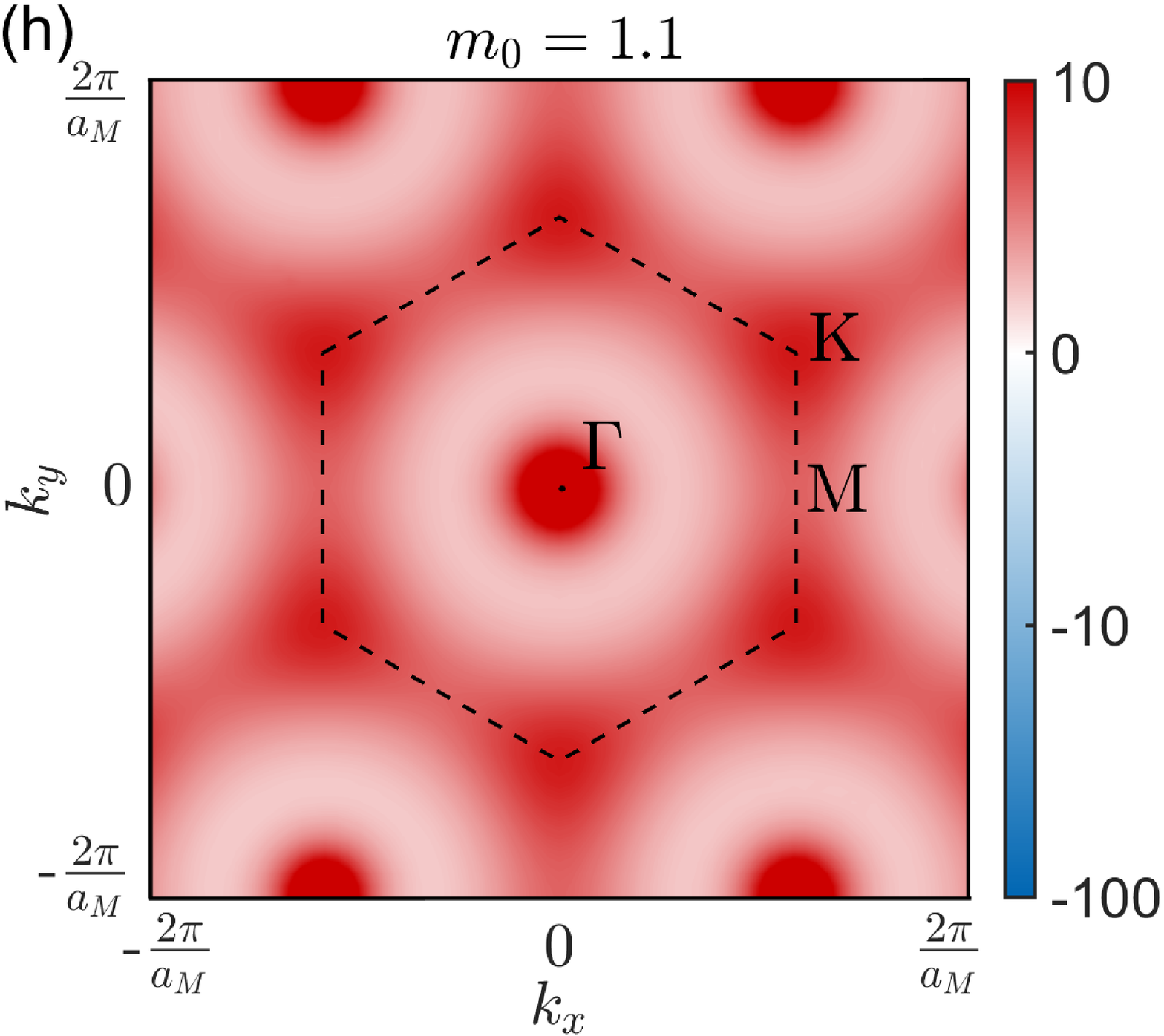}
    \end{tabular}
    \caption{(a)-(c) Moiré minibands of $H^\uparrow_{\mathbf k}(\mathbf r)$ for $a_M=5.7$ and (a) $m_0=-0.9$, (b) $m_0=1$ where a Dirac point appears and (c) $m_0=2.4$. The red dashed curves are band structures from the tight-binding model in Eq. \eqref{eq5}. Energies are in unit of $m_1$. Inset: heatmaps of $M(\mathbf r)$ characterizing the TMPs. (d) and (h): Berry curvature for the lowest band for $a_M=5.7$ and (d) $m_0=0.9$, (h) $m_0=1.1$. (e)-(g) Bloch function density $\rho(\mathbf r)$ (blue background) and current density $\mathbf j(\mathbf r)$ (red arrow) in real space for the top occupied band  at the $\Gamma$ point. The black solid curves are zero contour lines of $M(\mathbf r)$ as the topological domain walls. }
    \label{fig2}
\end{figure}

\section{Results}
\subsection{$C_{6z}$-symmetric domain walls} First we consider $C_{6z}$-symmetric domain walls, in which $\phi=0$, $m_0$ and $a_M$ are two tunable parameters, and mainly focus on electronic structures of $H^\uparrow_{\mathbf k}(\mathbf r)$. As shown in Fig. \ref{fig2}(a)-(c), for $a_M=5.7$, as $m_0<1$, all the low energy minibands are flat but topologically trivial. A gapped Dirac cone emerges at the CNP, arising from the closed domain-wall geometries \cite{ref13,ref56}. As $m_0=1$ a Dirac cone is formed at the $\Gamma$ point. As $m_0>1$, the two minibands around the CNP are inverted with spin Chern number $\pm1$, implying a topological phase transition at the CNP. Note that as $m_0>1.38$, the second and the third minibands are also inverted at the $\Gamma$ point with spin Chern number $\pm1$, thus topological flat minibands come out. The spin Chern number for each band is defined as the integral of the Berry curvature, which can be expressed as
\begin{equation}\label{eq6} 
\begin{aligned}
       &\Omega_z=-\text{Im}\{\bra{\partial_{\mathbf{k}}u_{n\mathbf{k}}}\times\ket{\partial_{\mathbf{k}}u_{n\mathbf{k}}}\}_z,\\
       &C_n=\frac{1}{2\pi}\int_{\text{BZ}}d^2\mathbf{k}\Omega_z,
	\end{aligned}
\end{equation}
where $\ket{u_{n\mathbf{k}}}$ denotes the periodic part of Bloch functions. As shown in Fig. \ref{fig2}(d) and (h), a $\pi$ flux switches near the $\Gamma$ point between the lowest two bands. For non-inverted bands, a -$\pi$ flux from the $\Gamma$ point is cancelled by a background flux of $\pi$, while for inverted bands they carry the same sign and add up to $2\pi$ and therefore the spin Chern number equals to 1. While the Chern number for each band is clearly defined as an integer, the total Chern number or the Hall conductance of all occupied bands can be half-integer values \cite{ref3}. The reason is because the Dirac cone near the CNP only contributes half a quantum of Hall conductance, while another half-integer values from high-energy fermions is obscured by the finite truncation of plane-wave basis. However, the gap-closing transition at the CNP shall be interpreted as a global topological phase transition because the total Chern number of all occupied bands is changed by 1.

Furthermore, with the topological phase transition driven by $m_0$, the Bloch states are substantially changed. As shown in Fig. \ref{fig2}(e)-(g), the wave function density of the first occupied miniband at the $\Gamma$ point is localized around the zero-contour lines of $M(\mathbf r)$, signifying the expected periodic edge states on the domain walls. Besides, the corresponding current density $\mathbf j(\mathbf r)=-ev\psi^{\dagger}[\sigma_x\hat{\mathbf x}+\sigma_y\hat{\mathbf y}]\psi$ implies chiral currents along the domain walls. For small $m_0$, chiral currents circulate in the clockwise direction around the circular domain walls enclosing domains on the triangular lattice with periodicity $a_M$. As $m_0$ increases, the domain walls expand and approach one another until $m_0=1.5$. Beyond this point, they contract and enclose domains on the honeycomb lattice centered around the triangular lattice. As a result, chiral currents circulate in the counter-clockwise direction along the domain walls. Additionally, the time-reversal part $H_\downarrow(\mathbf k,\mathbf r)$ has degenerate energy bands associated with opposite spin Chern numbers as well as chiral currents, resulting one helical channel on the periodic domain walls. 

\begin{figure}[tb]
  \centering
  \setlength{\tabcolsep}{0pt}
    \begin{tabular}{l l}
      \includegraphics[width=0.25\textwidth,valign=t]{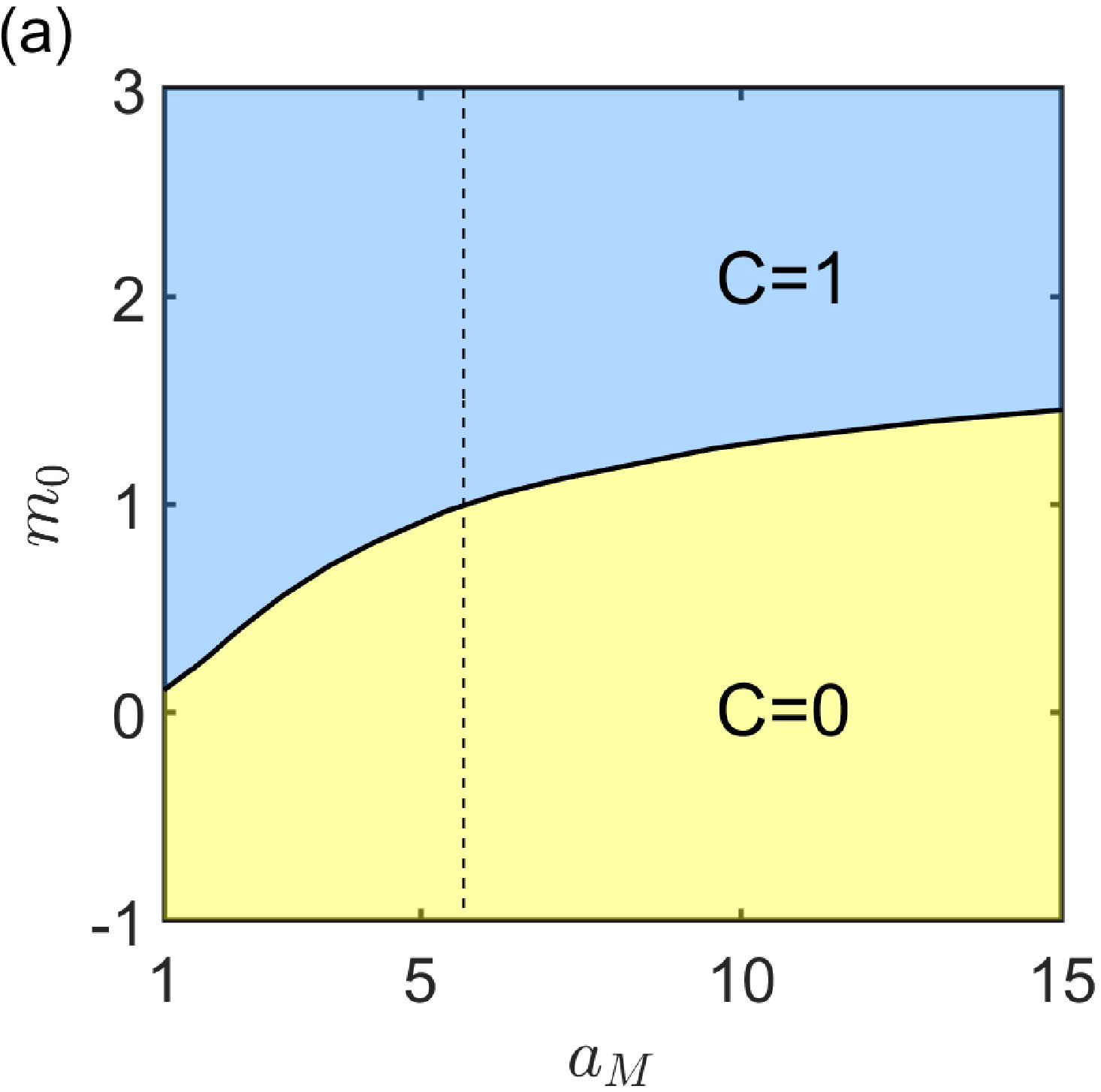} & 
      \includegraphics[width=0.24\textwidth,valign=t]{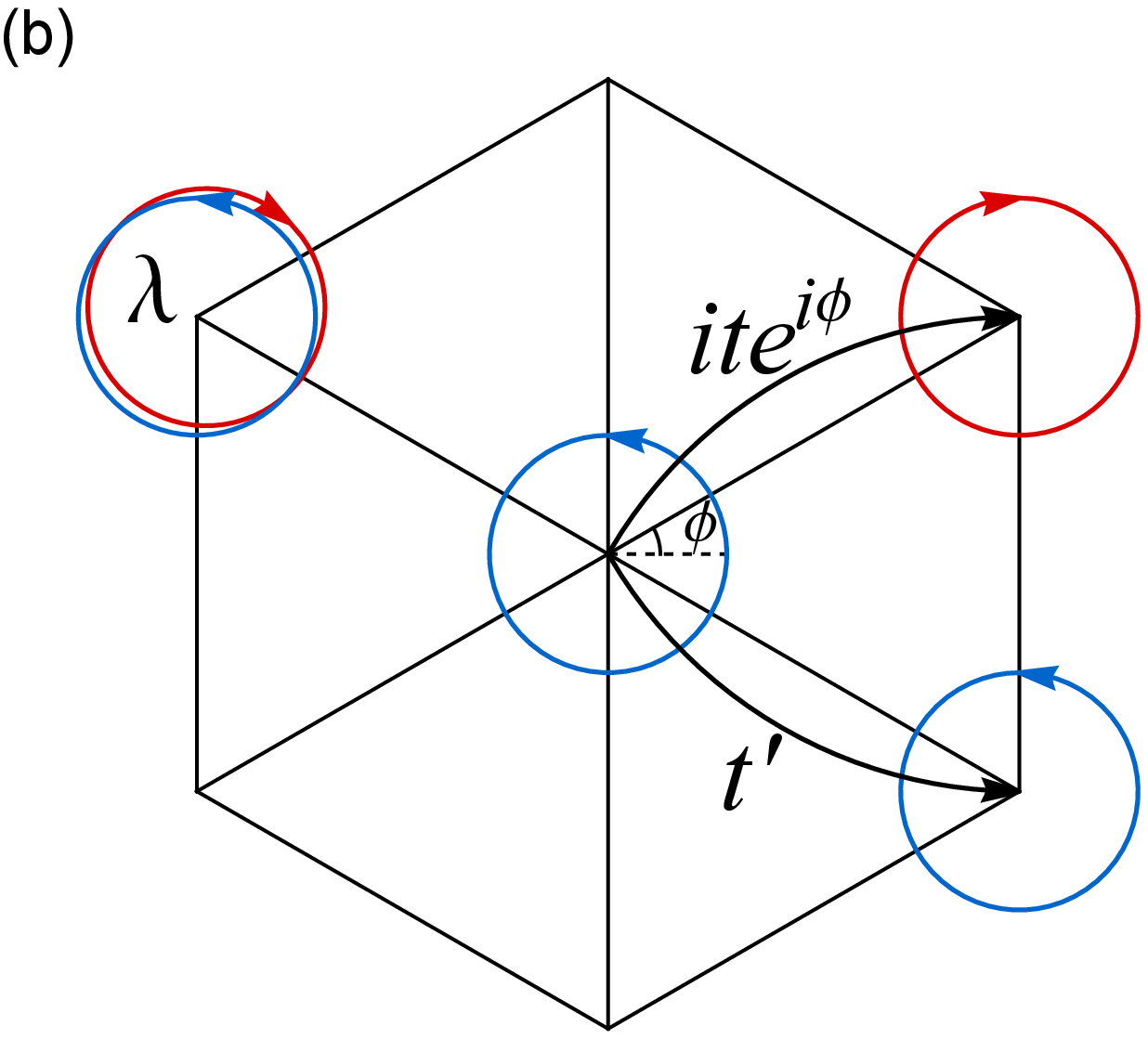} 
      \\[1.5\normalbaselineskip]
      \includegraphics[width=0.25\textwidth,valign=t]{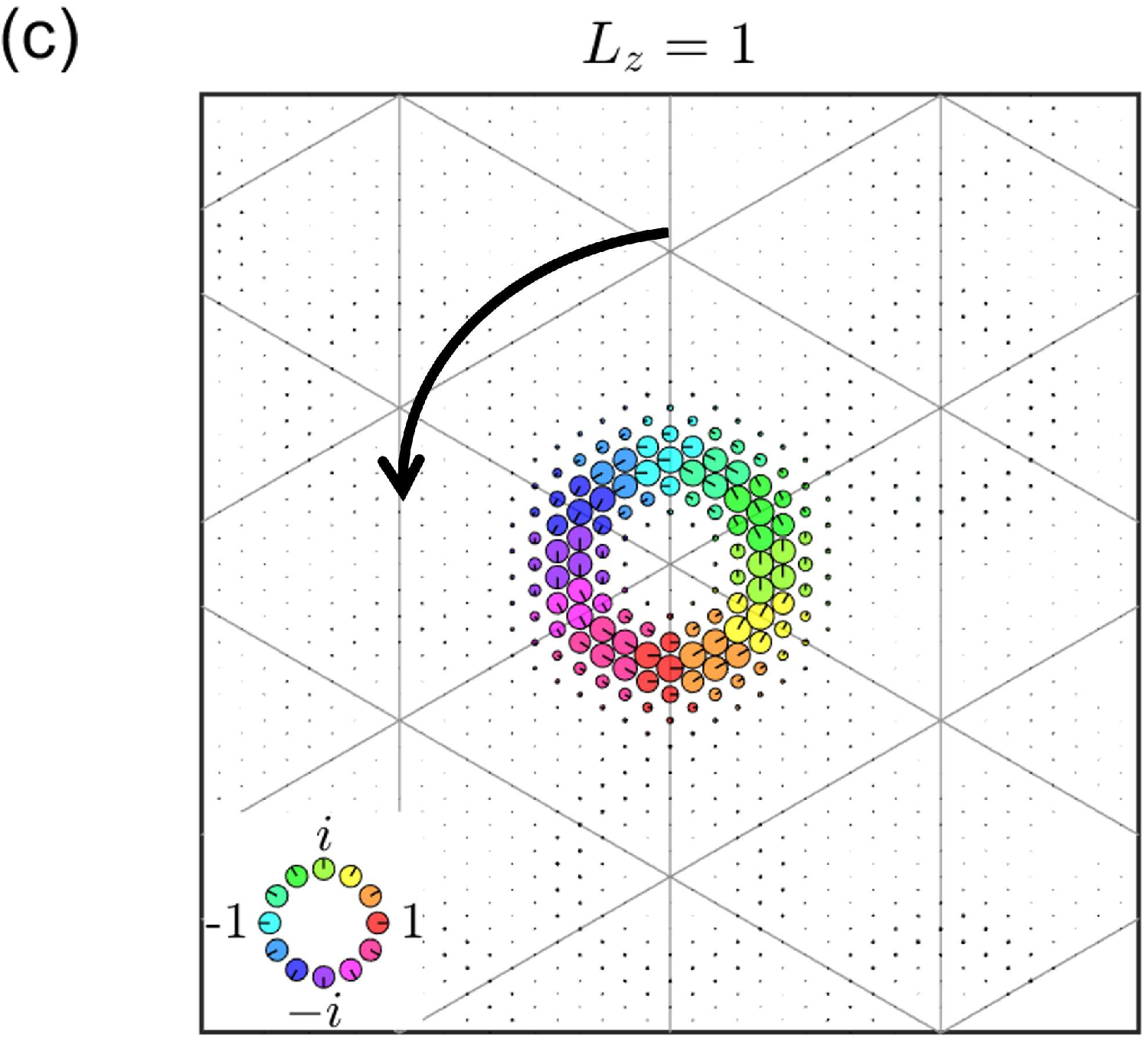} &
      \includegraphics[width=0.25\textwidth,valign=t]{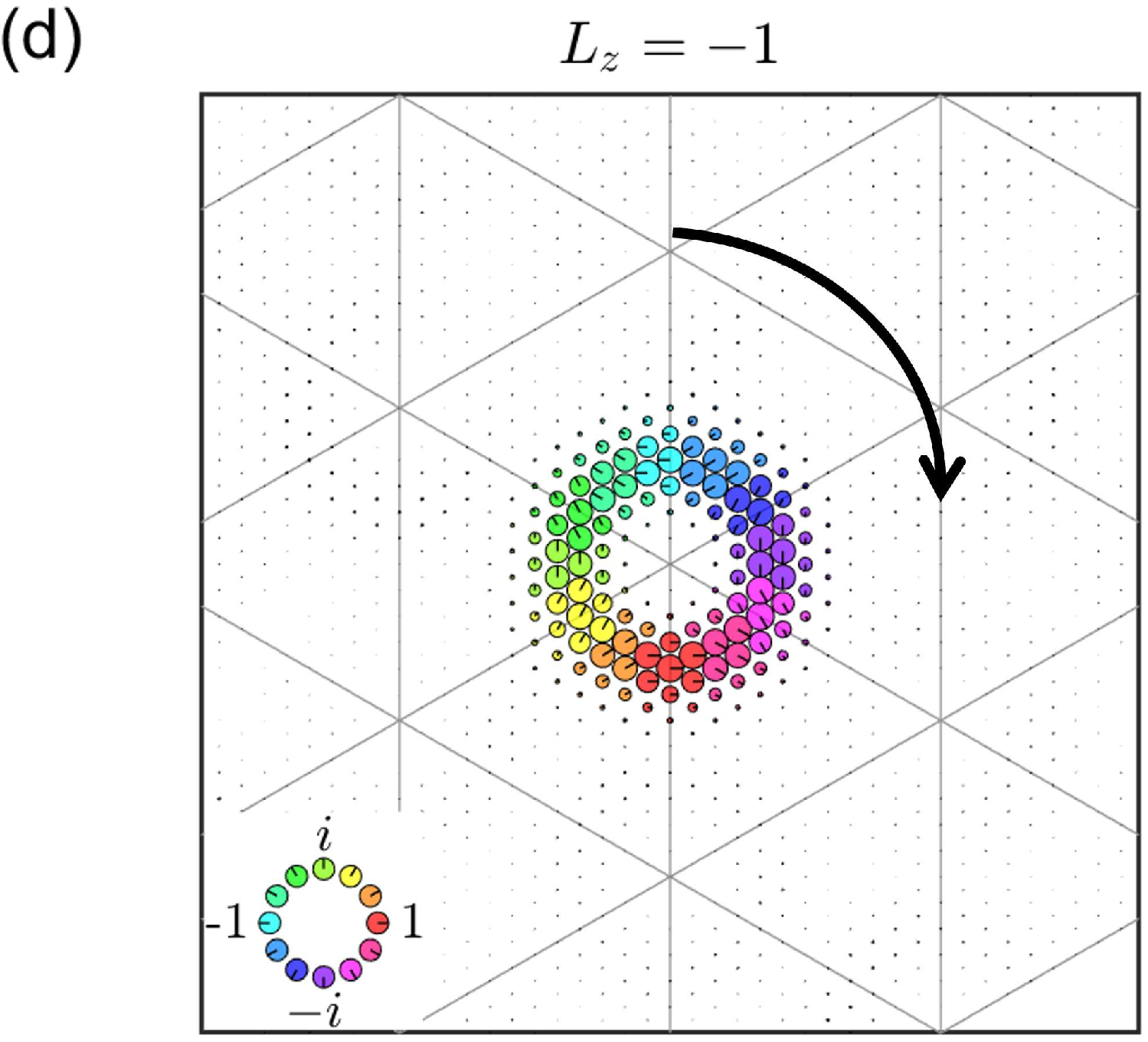} 
      \\[1.5\normalbaselineskip]
      \includegraphics[width=0.25\textwidth,valign=t]{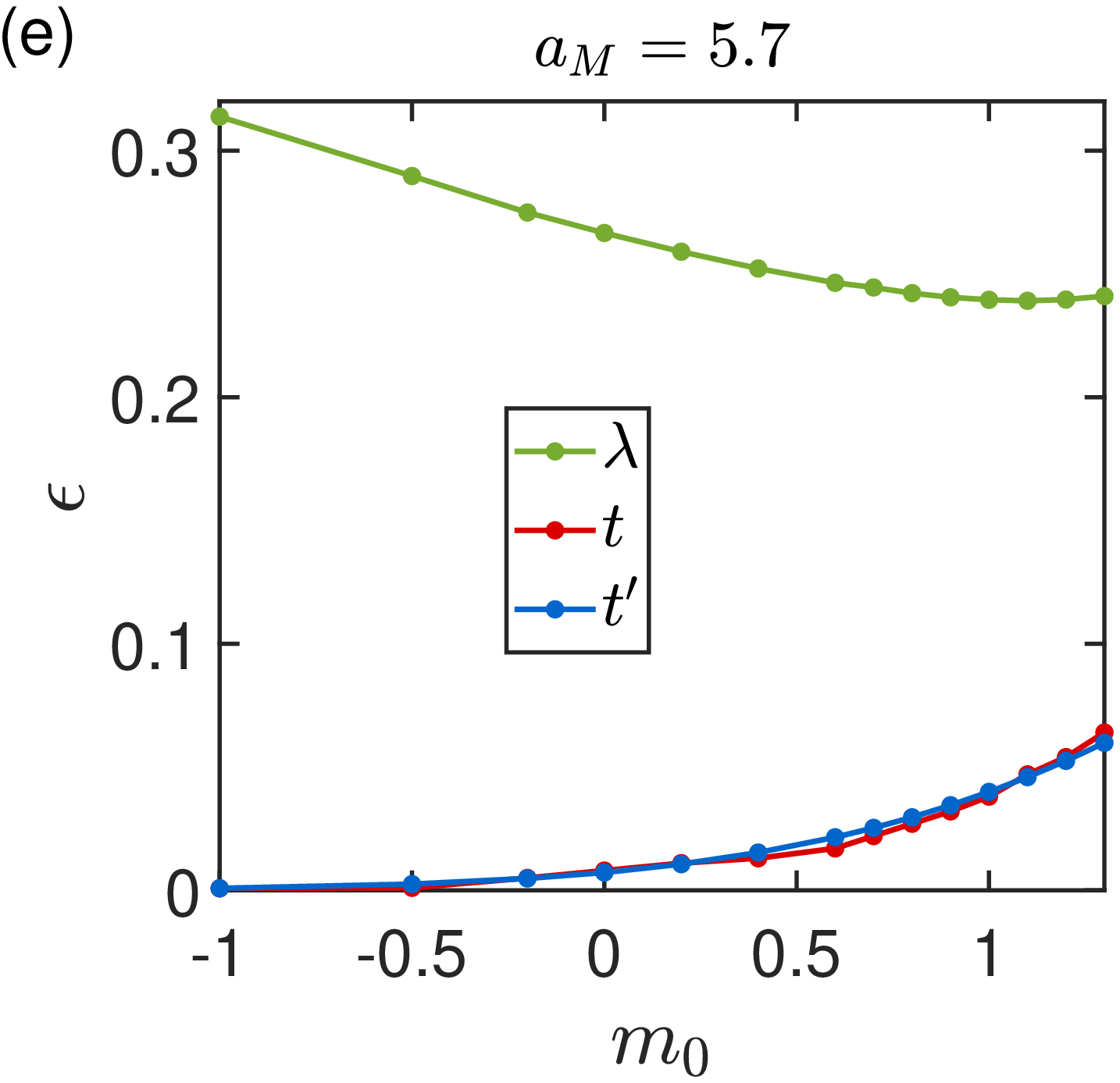} &
      \includegraphics[width=0.25\textwidth,valign=t]{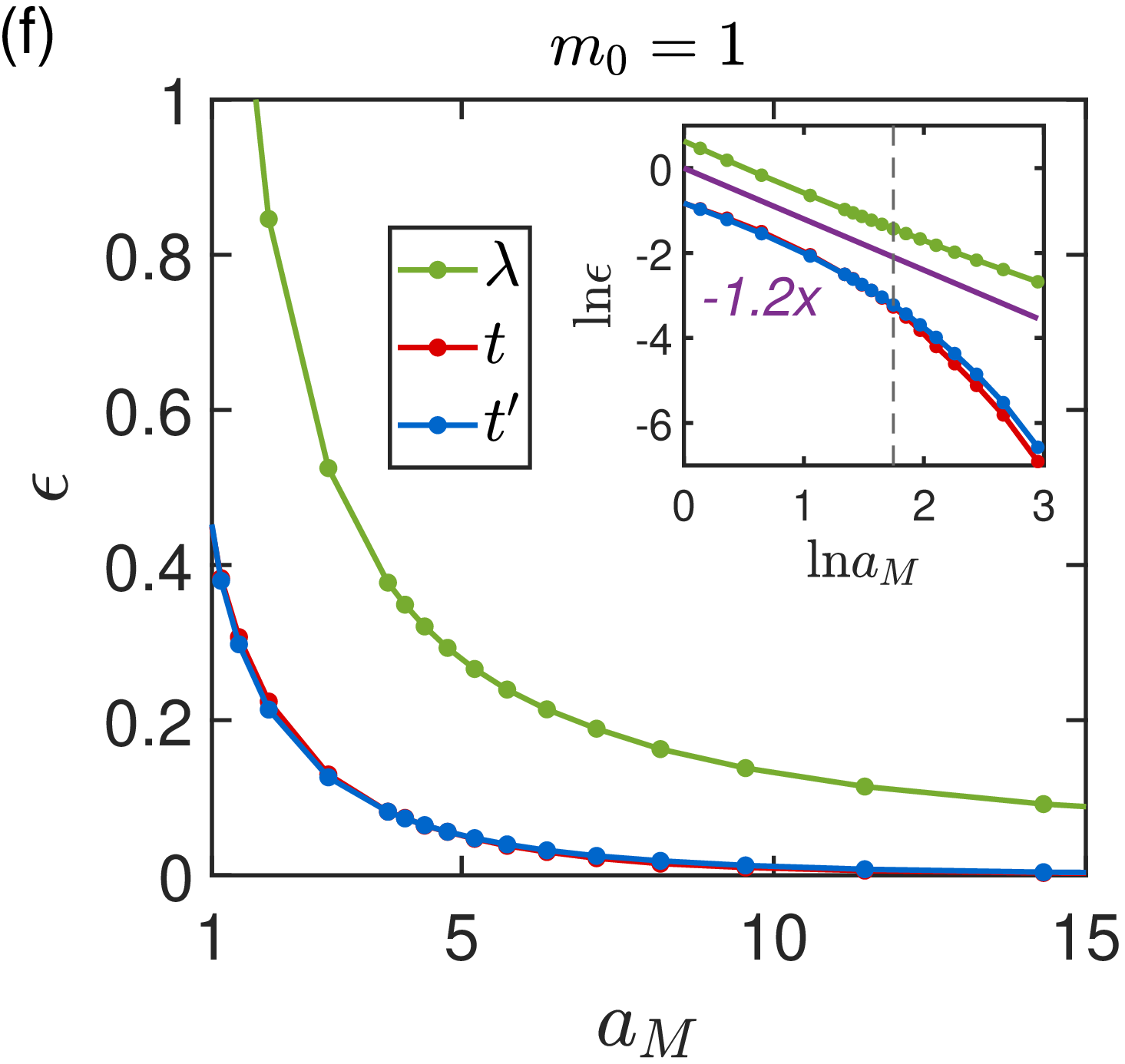} 
    \end{tabular}
    \caption{(a) Topological phase diagram of $C_{3z}$-symmetric domain walls tuned by (a) $a_M$ and $m_0$. The blue (yellow) regions represent two distinct topological phases with (without) band inversion at the CNP. The vertical dashed line contains parameters used in Fig. \ref{fig2}. (b) Sketch of the tight-binding model (5) on the triangular lattice. (c)-(d) Two Wannier functions associated with the two lowest minibands. Only one component of each Wannier spinor with non-vanishing orbital angular momentum is shown. (e) Hopping amplitudes as a function of $m_0$ for $a_M=5.7$. (f) Tight-binding parameters as a function of $a_M$ for $m_0=1$. Inset:
log-log plot of tight-binding parameters, where the dashed line denotes a critical point for $a_M$. All the tight-binding parameters have a power-law tail after the topological phase transition.}
\label{fig3} 
\end{figure}

In addition to $m_0$, we find the moiré periodicity $a_M$ could also drive the topological phase transition at the CNP, as illustrated in Fig. \ref{fig3}(a). Apparently this should be related to the interactions between periodic edge states. To gain deeper insight, we construct the maximally localized Wannier functions and a tight-binding model. We perform the standard maximally localized procedure to minimize the spread functional \cite{ref58}, with a proper gauge choice for the initial trial Wannier functions (see Appendix C). Interestingly, the localized Wannier orbitals for the lowest two minibands have characteristics of chiral $p_x\pm ip_y$ orbitals on the triangular lattice, as shown in Fig. \ref{fig3}(c)-(d). Up to the nearest neighbor hopping, the tight-binding model in the Wannier representations reads
\begin{equation}\label{eq5}
\begin{aligned}     
H_{\text{TB}}=&\lambda\sum_{i,\xi}\xi c^{\dagger}_{i\xi}c_{i\xi}+it\sum_{\langle i,j\rangle\xi}e^{-i\xi\phi_{ij}}c^{\dagger}_{i\xi}c_{j,-\xi}\\
-&t'\sum_{\langle i,j\rangle\xi}\xi c^{\dagger}_{i\xi}c_{j\xi},
	\end{aligned}
\end{equation}   
where $\xi=\pm$ and $c_{i,\pm}=(c_x\pm ic_y)/\sqrt2$ denotes the chiral basis associated with $p_{x}\pm ip_{y}$ orbitals. $\phi_{ij}$ is the azimuthal angle of the bond orientation from site $i$ to $j$. The first term denotes the on-site orbital rotation. The second and third terms represent the nearest-neighbor hopping between the orbitals with opposite and same orbital angular momentum, respectively, as shown in Fig. \ref{fig3}(b). The energy dispersion and band topology from Eq. \eqref{eq5} are both in good agreement with the lowest two minibands from Eq. \eqref{eq3} for various $m_0$ and $a_M$, as shown in Fig. \ref{fig2}(a)-(c). 

It is worth mentioning that this Hamiltonian is identical to the BHZ model on the triangular lattice \cite{ref74}. It gives rise to a Dirac cone at the $\Gamma$ point that can be expressed as $ H_{\text{eff}}=3a_Mt(k_x\sigma_x+k_y\sigma_y)+(\lambda-6t'+\frac32t'a_M^2k^2)\sigma_z$. For $\lambda<6t'$, the lower band carries Chern number +1, and under time reversal the chiral $p_{x}\pm ip_{y}$ orbitals are interchanged, $c_{\xi}\rightarrow c_{-\xi}$, so the lower band for the opposite spin part carries Chern number -1. Hence, the topological phase transition occurs as the hopping amplitudes $\lambda$,$t$, and $t'$ are changed by $m_0$ and $a_M$. The hopping amplitudes calculated from the constructed Wannier functions are shown in Fig. \ref{fig3}(e)-(f). For large $a_M$ or small $m_0$, the neighboring hopping between domain walls on the triangular lattice is very small. When increasing $m_0$ or decreasing $a_M$, the domain walls approach each other as seen in \ref{fig2}(e) and (f), and the chiral $p_{x}\pm ip_{y}$-orbital electrons on the domain walls are more likely to hop between the domain walls, namely, $t$ and $t'$ increase. Consequently, the bandwidth increases and the band gap decreases to facilitate the band inversion. It is noteworthy as $m_0>1.38$, the lowest four minibands can be described by the tight-binding model in Eq. \eqref{eq5} on the honeycomb lattice, and realize two copies of the Haldane model for each spin \cite{ref60}. Moreover, the on-site orbital rotation on the honeycomb lattice plays a similar role with the neighboring hopping on the triangular lattice (see Appendix D), which is crucial for topological phase transition in the TMP. After the topological phase transition, the hopping amplitudes have a power-law tail with respect to $a_M$ due to the topological obstruction \cite{ref57,ref59,ref75}, as shown in Fig. \ref{fig3}(f) inset.

\subsection{$C_{3z}$-symmetric domain walls} 
\begin{figure}[tb] 
  \centering
  \setlength{\tabcolsep}{0pt}
    \begin{tabular}{l l}
      \includegraphics[width=0.24\textwidth,valign=t]{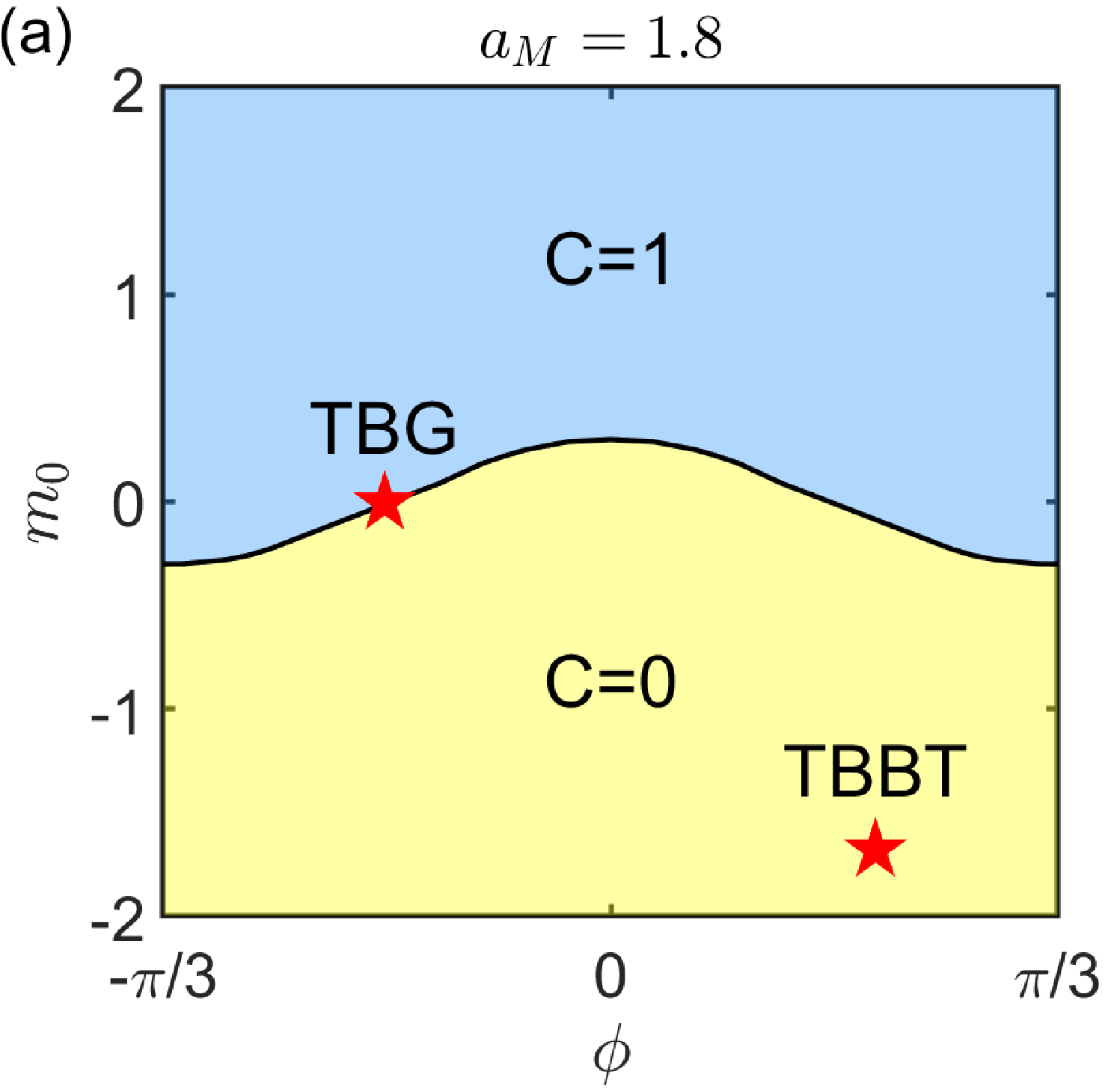} & 
      \includegraphics[width=0.24\textwidth,valign=t]{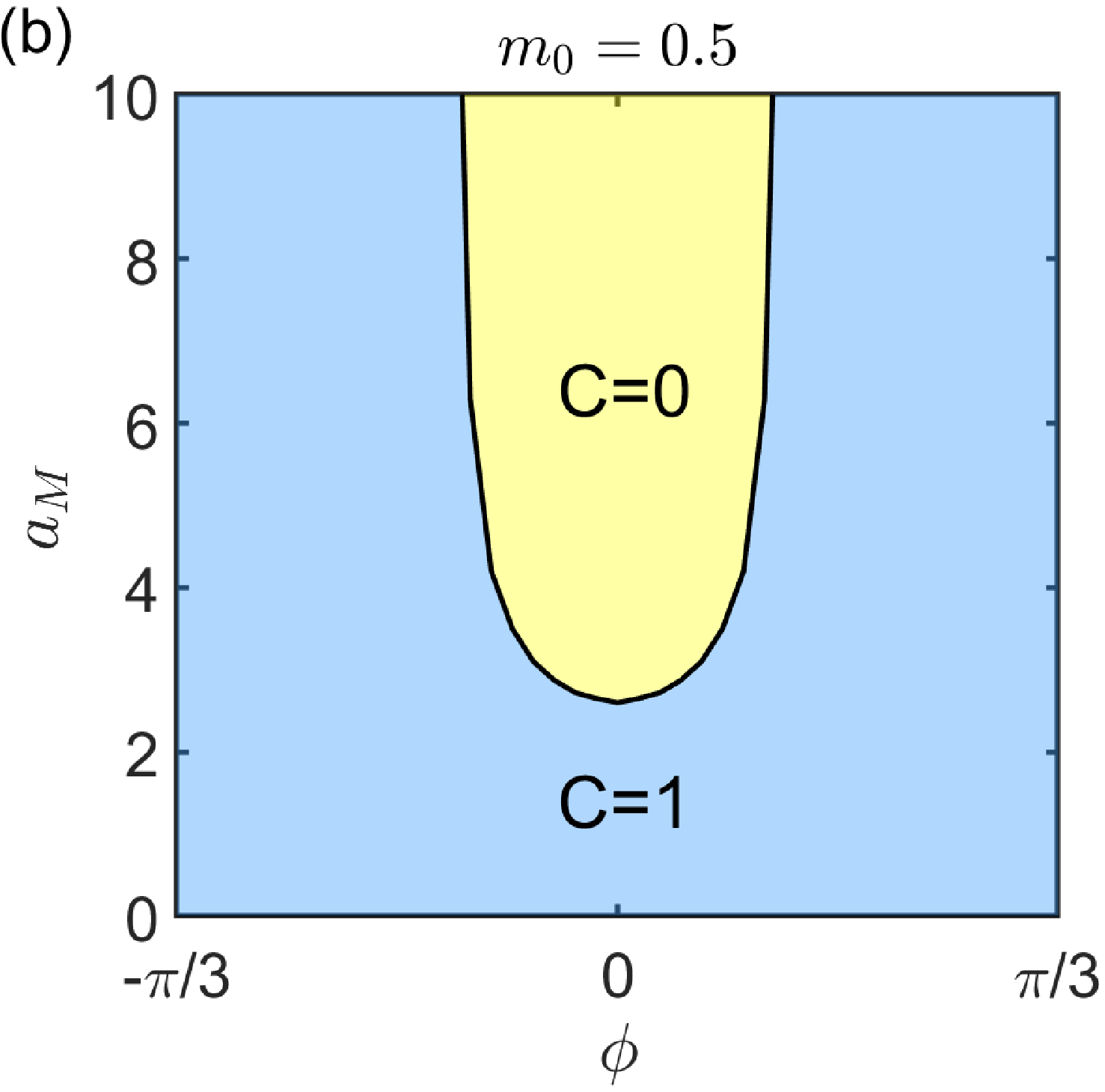} 
      \\[1.5\normalbaselineskip]
      \includegraphics[width=0.24\textwidth,valign=t]{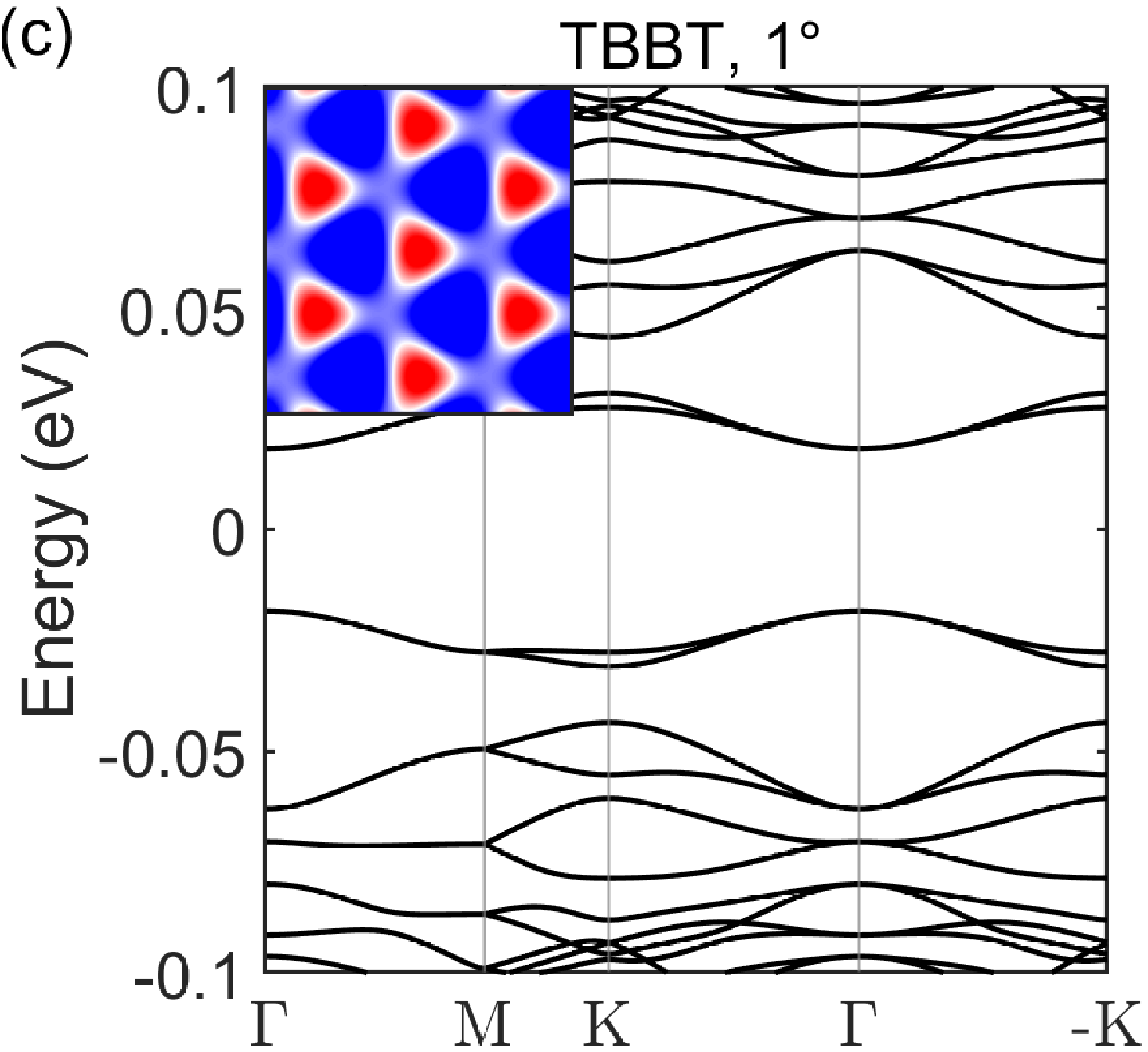} &
      \includegraphics[width=0.24\textwidth,valign=t]{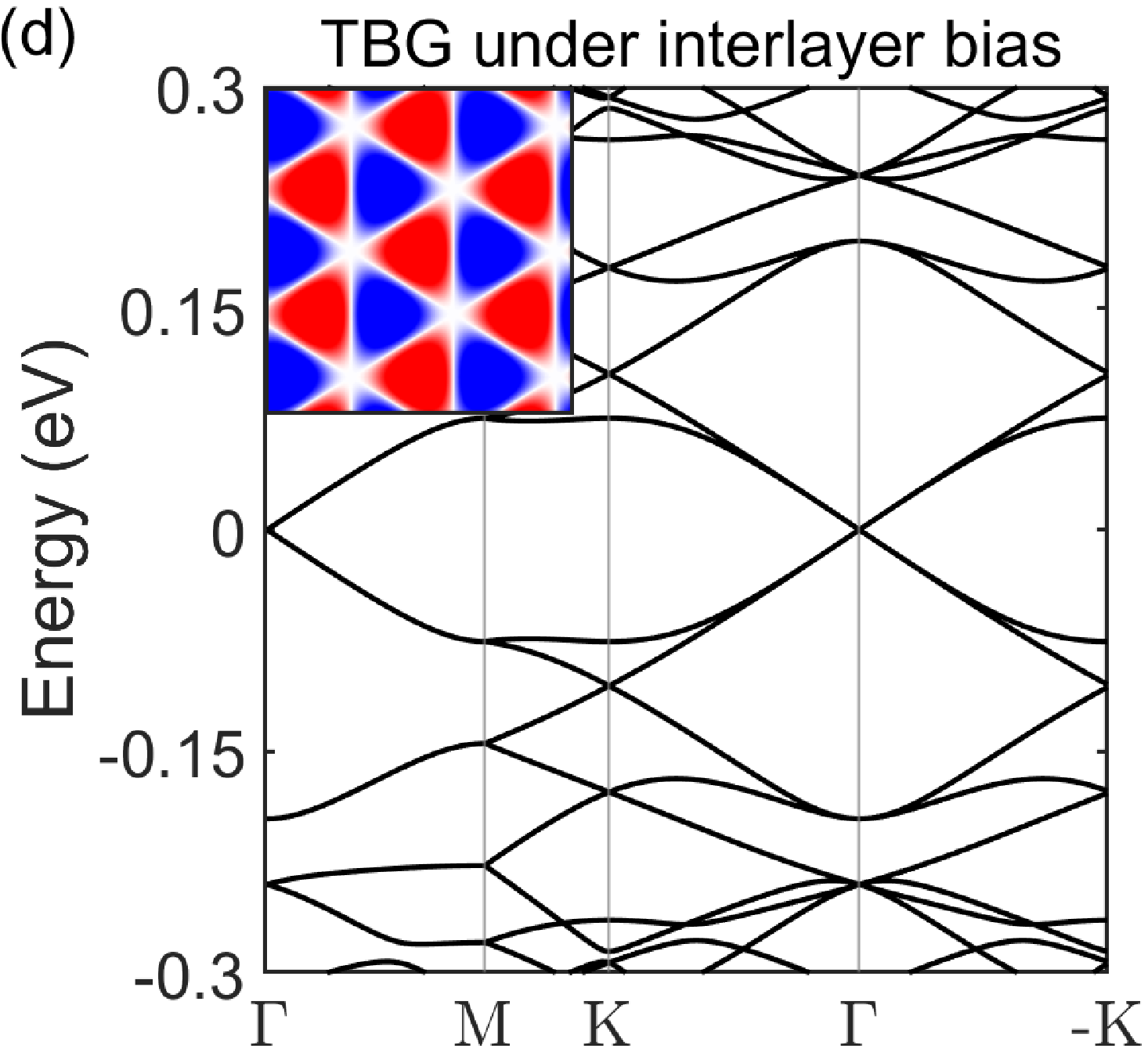}  
    \end{tabular}
    \caption{(a)-(b) Topological phase diagram of $C_{3z}$-symmetric domain walls tuned by (a) $\phi$ and $m_0$ for $a_M=1.8$, and (b) $\phi$ and $a_M$ for $m_0=0.5$. The blue (yellow) regions represent two distinct topological phases with (without) band inversion at the CNP. The red stars mark topological phases of TBBT and domain wall network of TBG. (c)-(d) Moiré minibands of (c) TBBT at $1^{\circ}$ twist angle, and (d) domain wall network of TBG under interlayer bias, the parameters are calculated from the band gaps for local registries \cite{ref13,ref55}. Inset: heatmaps of $M(\mathbf r)$ characterizing the TMPs.}
\label{fig4} 
\end{figure}

Most moiré systems have moiré superlattices that are symmetric under $C_{3z}$ operations. This makes it more practical to study the topological phase transition (TMP) with $C_{3z}$-symmetric domain walls, which involves another parameter $\phi$. When $\phi$ deviates from $2n\pi/3$, it breaks the inversion symmetry and causes the shapes of domain walls to change. When $\phi$ approaches $\pi/6$, a triangular domain wall network will be formed. This change has a minor effect on the domain wall size when compared to the effect of $m_0$, since $m_0$ can always drive the topological phase transition whatever $\phi$ is, as illustrated in Fig. \ref{fig4}(a). However, as $m_0$ is tuned around the phase boundary, $\phi$ and $a_M$ can also drive the topological phase transition, as illustrated in Fig. \ref{fig4}(b). To be specific, $a_M$ will significantly close and reopen the miniband gap only if $\phi$ is close to $2n\pi/3$, as shown in Fig. \ref{fig4}(b).

To demonstrate the versatility of our model, we take TBBT and TBG under interlayer bias as two practical paradigms, which are proposed to exhibit isolated circular shapes and connected triangular shapes of $C_{3z}$-symmetric domain walls \cite{ref13,ref54,ref55,ref73}, respectively. Take TBBT as an example, $a_M\approx a_0/\theta$ is about 25 nm at $1^{\circ}$ twist angle, and $v\approx250$ $\text{meV}\cdot\text{nm}$ \cite{ref20,ref21}. We take $(m_0,m_1,\phi)$ as (-31meV,18meV,0.62) that can be estimated from the band gaps of untwisted high-symmetry atomic registries \cite{ref13}, so $m_0/m_1\approx-1.7$ and $a_Mm_1/v\approx1.8$, marked in Fig. \ref{fig4}(a). The band structure of TBBT from those parameters exhibits a miniband gap at the CNP, as shown in Fig. \ref{fig4}(c), and the two lowest bands remain non-inverted. It is important to note that the moiré minibands derived from the Dirac Hamiltonian in Eq. \eqref{eq1} represent only a small fraction of the TBBT bands, so our model cannot determine the Hall conductance or the global TI at the CNP as mentioned earlier. Nevertheless, the low-energy miniband dispersion and topology align well with prior research \cite{ref13}. As a result, a global TI/NI phase transition could occur if the domain walls are adjusted to close and reopen the miniband gap at the CNP. This may be achievable by considering lattice reconstruction, as TI domains are expected to enlarge due to local atomic registries with an inverted band regime being more stable than other atomic registries \cite{ref13}. Consequently, the CNP gap could decrease or even undergo a topological phase transition in a fully relaxed TBBT system.

On the other hand, the domain wall network in twisted bilayer graphene (TBG) features two gapless Dirac cones, as illustrated in Fig. \ref{fig4}(d), resulting in a metallic system. Our model, presented in Eq. \eqref{eq1}, accurately reproduces the band structure using a simple tight-binding model for the domain wall network. However, as our model only accounts for a single helical channel, the local valley Chern number across TBG network domain walls is 2 \cite{ref54}, allowing for two helical channels per valley. Consequently, the coupling between these channels within the AA region is absent, a factor that has been suggested as critical for Aharonov–Bohm oscillations observed in transport experiments \cite{ref77,ref78,ref79}. A more precise model incorporating the coupling between the two channels can be further developed.

\section{Discussion} So far we have unveiled that the TMP with $C_{6z}$-symmetric domain walls has chiral $p_x\pm ip_y$ orbitals in the low energy minibands. In our model, we neglect the particle-hole asymmetric part that will bring about a scalar moiré potential that involves more complicated topological phase diagrams \cite {ref14}. While the continuum model in Eq. \eqref{eq1} works well for $C_{3z}$-symmetric domain walls, the tight-binding model in Eq. \eqref{eq5} can also be generalized by adding a sublattice asymmetry term on the honeycomb lattice \cite{ref30}. In general, a TMP can generate chiral $p$-orbital bands with nontrivial topology.

Although the $p_x\pm ip_y$-orbital model on the honeycomb lattice has been seen in low energy minibands in TBG \cite{ref68,ref69}, the on-site orbital rotation is negligible to induce a sizeable topological miniband gap at the CNP. On the other hand, the geometry of domain walls in TMP can be tuned by twist angle, interlayer bias, and lattice strain, so the electron states are allowed to be more spatially extended with certain chirality, which can lead to topological flat bands as discussed above. Hence the chiral $p_x\pm ip_y$ orbitals on the domain walls may bring strong correlation effect such as Wigner crystal and ferromagnetism \cite{ref52,ref66,ref67}. The periodic edge states will be expected to involve anisotropic $p$-band interactions which are rarely found in realistic materials, while isotropic $s$-band interactions have been proposed in twisted TMDs materials \cite{ref46,ref61,ref62,ref63,ref64}.

\begin{acknowledgments} H. Wang thanks Yafei Ren, Changpeng Lin and Shang Ren for invaluable discussions. H.W. is supported by the Air Force Office of Scientific Research (AFOSR) Grant No. FA9550-20-1-0255, and L.Y. is supported by the National Science Foundation (NSF) Grant No. DMR-2124934.
\end{acknowledgments}

\appendix
\section{Underlying Dirac equation from twisted bilayer BHZ model}
In this section we display the relationship between our modified Dirac Hamiltonian and the twisted bilayer BHZ Hamiltonian proposed by Lian \textit{et al.} to describe twisted bilayer MnBi$_2$Te$_4$ moiré superlattice \cite{ref12}. This layered system has two distinct magnetic orders, ferromagnetic/antiferromagnetic, whose Hamiltonian is different from each other. Even so, it has a non-magnetic part which is same for both. It consists of a two-by-two block matrix, where the intralayer (diagonal) blocks with a relative twist angle are coupled via interlayer (off-diagonal) blocks. The non-magnetic part reads
\begin{equation}
\mathcal{H}_1=\begin{pmatrix}
    h_{\theta/2}(-i\nabla) & T(\mathbf r) \\
    T^{\dagger}(\mathbf r) & h_{-\theta/2}(-i\nabla)
\end{pmatrix},
\end{equation}
where $h_{\pm\theta/2}=R^{\dagger}_{\pm\theta/2}\mathcal{H}_{\text{BHZ}}R_{\pm\theta/2}$ describes the intralayer coupling. Each layer is rotated by $\pm\theta/2$ about $z$-axis so the relative twist angle between two layers is $\theta$. In basis of $\{|\text{Bi}^\nu,\uparrow\rangle,|\text{Te}^\nu,\downarrow\rangle,|\text{Bi}^\nu,\downarrow\rangle,|\text{Te}^\nu,\uparrow\rangle\}$ where $|\text{Bi}^\nu,s\rangle$ is $\nu$-th layer $p_z$ orbital of Bi atoms with spin $s$, the rotation matrix $R_{\pm\theta/2}$ can be expressed as $\text{diag}(e^{\pm i\theta/4},e^{\mp i\theta/4},e^{\pm i\theta/4},e^{\mp i\theta/4})$ and 
\begin{equation}
\mathcal{H}_{\text{BHZ}}=\epsilon(\mathbf k)+\begin{pmatrix}
    m(\mathbf k) & \alpha k_{-} & & \\
   \alpha k_{+} & -m(\mathbf k) & &\\
    & & m(\mathbf k) & \alpha k_{+}\\
    & &\alpha k_{-} & -m(\mathbf k)\\
\end{pmatrix},
\end{equation}
\begin{equation}
    T(\mathbf r)=\begin{pmatrix}
    t_{11} & & & it_{12} \\
     & -t_{22} & -it_{12}&\\
    & it_{12}& t_{11} & \\
    -it_{12}& & & -t_{22}\\
\end{pmatrix},
\end{equation}
where $T(\mathbf r)$ describes the interlayer coupling. With the relative layer displacement slowly varying over the moiré periodicity, the interlayer coupling for local atomic registries are continuously tuned. Hence $T(\mathbf r)$ is real position dependent with moiré periodicity. This block matrix can be partially decoupled by introducing a unitary transformation for the basis, 
\begin{equation}
\begin{aligned}
    &|\text{Bi}^\pm,s\rangle=\frac{1}{\sqrt{2}}\Big(|\text{Bi}^1,s\rangle\mp|\text{Bi}^2,s\rangle\Big),\\
    &|\text{Te}^\pm,s\rangle=\frac{1}{\sqrt{2}}\Big(|\text{Te}^1,s\rangle\mp|\text{Te}^2,s\rangle\Big),
	\end{aligned}
\end{equation}
where $|\text{Bi}^+,s\rangle$ is the bonding state of $p_z$ orbital of Bi atoms of two twisted layers with spin $s=\uparrow,\downarrow$, and $|\text{Bi}^-,s\rangle$ is the antibonding state of $p_z$ orbital of Bi atoms of two twisted layers with spin $s$. This unitary transformation mixes states in the two twisted layers. After the unitary transformation the Hamiltonian can be written as
\begin{widetext}
\begin{equation}
\begin{aligned}
	&\mathcal{H}_1=\epsilon(\mathbf k)\bold{1}_{8\times8}+\\
   &\begin{pmatrix}
   m(\mathbf k)+t_{11} & \alpha\cos{\frac{\theta}{2}}k_{-} & & it_{12} & & -i\alpha\sin{\frac{\theta}{2}} k_{-}& \\
   \alpha\cos{\frac{\theta}{2}}k_{+} & -m(\mathbf k)-t_{22} & -it_{12} & &-i\alpha\sin{\frac{\theta}{2}}k_{+}\\
    & it_{12} & m(\mathbf k)+t_{11} & \alpha\cos{\frac{\theta}{2}}k_{+} &&&&-i\alpha\sin{\frac{\theta}{2}}k_{+}\\
    -it_{12} & &\alpha\cos{\frac{\theta}{2}}k_{-} & -m(\mathbf k)-t_{22} &&&-i\alpha\sin{\frac{\theta}{2}}k_{-}\\
    &i\alpha\sin{\frac{\theta}{2}}k_{-}&&&m(\mathbf k)-t_{11}&-\alpha\cos{\frac{\theta}{2}} k_+ & & it_{12}\\
    i\alpha\sin{\frac{\theta}{2}}k_{+}&&&&-\alpha\cos{\frac{\theta}{2}} k_-&-m(\mathbf k)-t_{22}&-it_{12}&\\
    &&&i\alpha\sin{\frac{\theta}{2}}k_{+}&&it_{12}&-m(\mathbf k)-t_{11}&-\alpha\cos{\frac{\theta}{2}} k_-\\
    &&i\alpha\sin{\frac{\theta}{2}}k_{-}&&-it_{12}&&-\alpha\cos{\frac{\theta}{2}} k_+&m(\mathbf k)-t_{22}
\end{pmatrix},
\end{aligned}
\end{equation}
\end{widetext}
where $t_{11}$ and $t_{22}$ are positive numbers representing the interlayer hopping between Bi-Bi and Te-Te atoms, respectively. Then the Hamiltonian appears to be block diagonal with the zeroth-order approximation at small twist angle where $\sin\frac\theta2\ll\cos\frac\theta2$. The top left block now reads
\begin{widetext}
\begin{equation}
\begin{aligned}
\mathcal{H}_2=\epsilon(\mathbf k)\bold{1}_{4\times4}+\begin{pmatrix}
   m(\mathbf k)+t_{11} & \alpha k_{-} & & it_{12} \\
   \alpha k_{+} & -m(\mathbf k)-t_{22} & -it_{12} & \\
    & it_{12} & m(\mathbf k)+t_{11} & \alpha k_{+} \\
    -it_{12} & &\alpha k_{-} & -m(\mathbf k)-t_{22} \\
\end{pmatrix},
\end{aligned}
\end{equation}
\end{widetext}
in the basis of $\{|\text{Bi}^-,\uparrow\rangle,|\text{Te}^+,\downarrow\rangle,|\text{Bi}^-,\downarrow\rangle\,|\text{Te}^+,\uparrow\rangle\}$. This decoupled matrix is time-reversal invariant and has degenerate eigenvalues $\delta\pm\sqrt{M^2+k^2}$ where $\delta=\epsilon(\mathbf k)+\frac12\big(t_{11}-t_{22}\big)$ and $M=\sqrt{t_{12}^2+\Big(m+\frac{t_{11}+t_{22}}{2}\Big)^2}$. Hence the Hamiltonian can be transformed without breaking time-reversal symmetry as
\begin{equation}
\mathcal{H}_2=\delta\bold1_{4\times4}+\begin{pmatrix}
     \mathbf d\cdot\boldsymbol\sigma & \\
     & (\mathbf d\cdot\boldsymbol\sigma)^{\dagger}
\end{pmatrix},
\end{equation}
where $\mathbf d=(\alpha k_x,\alpha k_y,M)$ and $\boldsymbol\sigma$ are the Pauli matrice. Apparently $M=M(\mathbf r)$ has the moiré periodicity that is dependent on twist angle. Therefore the particle-hole symmetric part of the Hamiltonian is mathematically equivalent to the modified Dirac Hamiltonian in the main text. Apparently the interlayer hopping parameters play a role as the periodic modulation on the mass term, which is also reflected by the fact that the interlayer hopping could alter topological properties of the local atomic registries and thus generate the TMP in moiré superlattices.

Besides, we note that our modified Dirac Hamiltonian in Eq. \eqref{eq1} in the main text is identical to the twisted BHZ model with a moiré scale oscillation in the mass term \cite{ref13} (which can be simplified from the twisted bilayer BHZ model as mentioned above) to describe twisted bilayer Bi$_2$Te$_3$. This also makes sense since the original BHZ model itself can be used to describe the thin films as well as bulk of topological insulators \cite{ref25}. Even so, these two versions of BHZ model may have different physical interpretations in the sense that the basis has been changed under unitary transformation.

\section{Derivation of spin-orbit coupling in modified Dirac Hamiltonian}
In this section we provide a further evaluation on the modified Dirac Hamiltonian to show the intrinsic spin-orbit coupling (SOC), from which we may have partial insight into the origin of chiral $p_x\pm ip_y$-orbitals for the low energy bands. Suppose the eigenvector for  $H^\uparrow_{\mathbf k}(\mathbf r)$ is $(\psi_1,\psi_2)^T$, then the corresponding eigenvalue equation can be written as
\begin{equation}
\begin{pmatrix}
     M & k_x-ik_y\\
     k_x+ik_y & M
\end{pmatrix}\begin{pmatrix}
     \psi_1\\
     \psi_2
\end{pmatrix}
=\epsilon\begin{pmatrix}
     \psi_1\\
     \psi_2
\end{pmatrix},
\end{equation}
which involve two coupled equations. From the second equation we obtain $\psi_2=(\epsilon+M)^{-1}(k_x+ik_y)\psi_1$. By substituting $\psi_2$ in the first equation and using the canonical commutation relation the eigenvalue equation for $\psi_1$ can be expressed as  
\begin{equation}
\begin{aligned}
        \bigg(-\nabla^2+M^2+i\frac{1}{M}
        \big(\nabla M&\times\nabla\big)_z+\\
        &\frac{1}{M}\nabla M\cdot\nabla\bigg)\psi_1=\epsilon^2\psi_1,
	\end{aligned}
\end{equation}
where we have used the low-energy approximation $\frac{1}{M+\epsilon}\approx\frac1M$. Combining spin and orbital pseudospin degrees of freedom, the total eigenvalue equation is decoupled as
\begin{equation}
\begin{aligned}
        \bigg(-\nabla^2+V+i s_z\sigma_z\big(\frac{1}{2V}&{\nabla V\times\nabla}\big)_z+\\
        &\frac{1}{2V}\nabla V\cdot\nabla\bigg)\Psi=\epsilon^2\Psi,
	\end{aligned}
\end{equation}
where $V=M(\mathbf r)^2$. Now we arrive at a relativistic wave equation. The first two terms are kinetic and potential energy that generate bound states around the potential minimum. The third term denotes the atomic SOC that diverges as $M(\mathbf r)$ reaches 0, corresponding to the domain walls in TMP. The atomic SOC from the modified Dirac Hamiltonian possibly leads to the chiral $p_x\pm ip_y$ orbitals per spin in low energy and thus non-trivial band topology.

\begin{figure*}[tb]
  \centering
      \includegraphics[width=\textwidth]{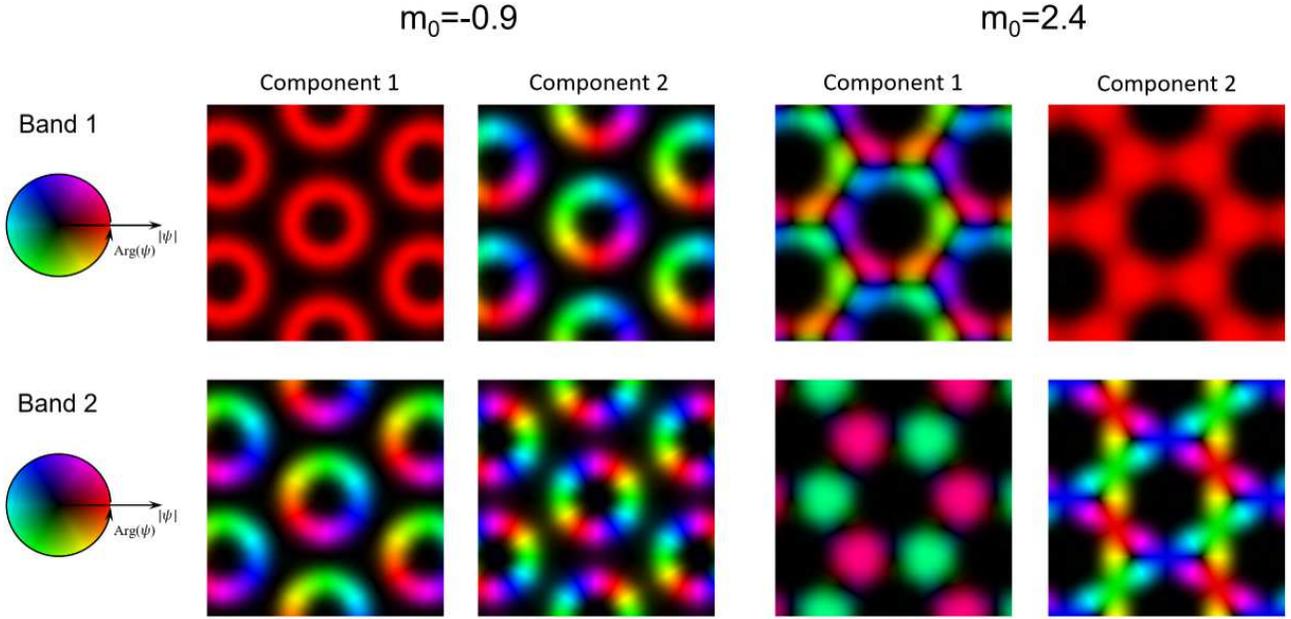} 
    \caption{Real space distributions of the Bloch spinors at the $\Gamma$ point for the top two occupied bands. The bottom two unoccupied bands are related by the particle-hole symmetry.}
    \label{fig6}
\end{figure*}
\begin{figure*}[tb]
  \centering
      \includegraphics[width=0.24\textwidth,valign=t]{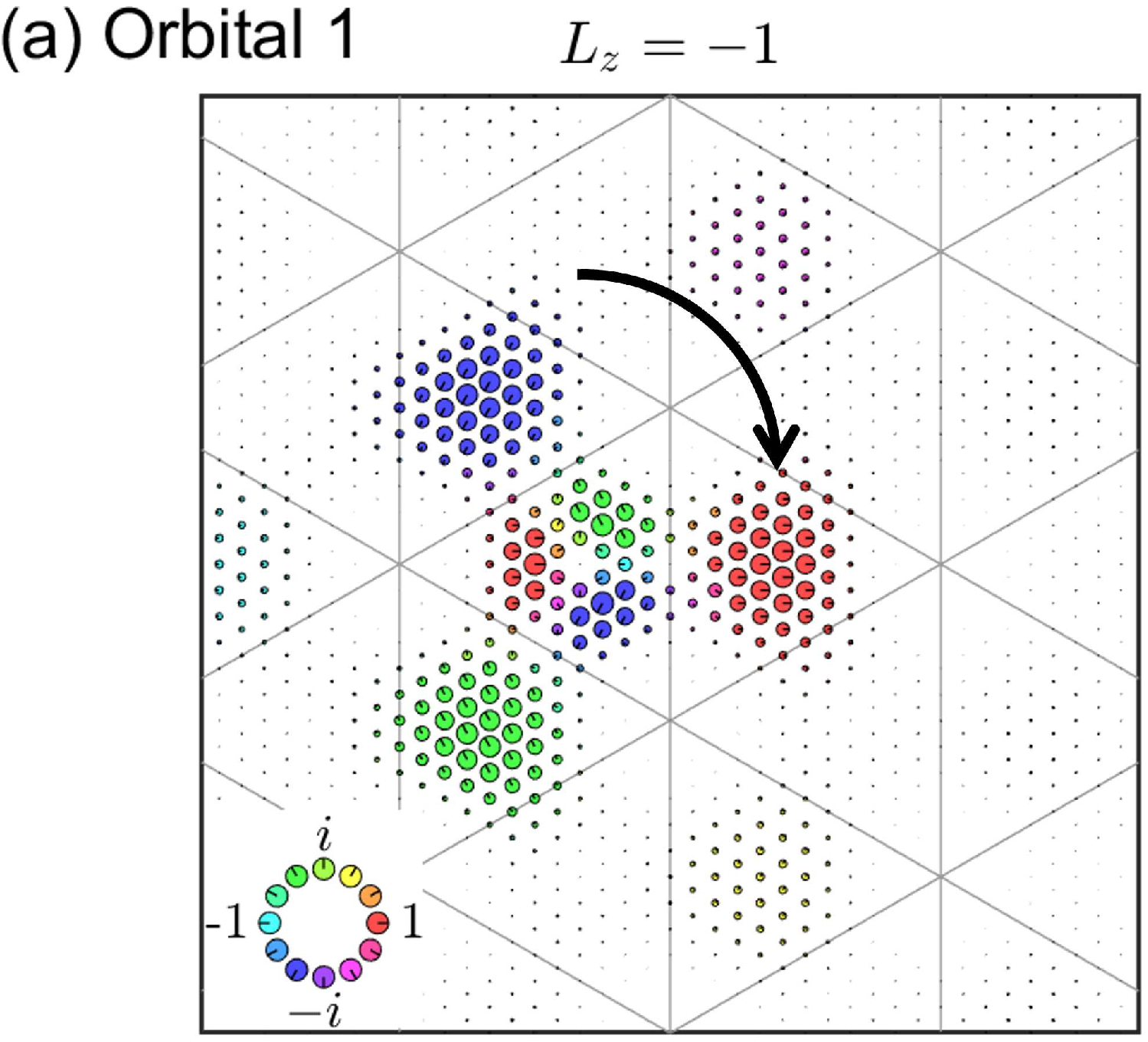} 
      \includegraphics[width=0.24\textwidth,valign=t]{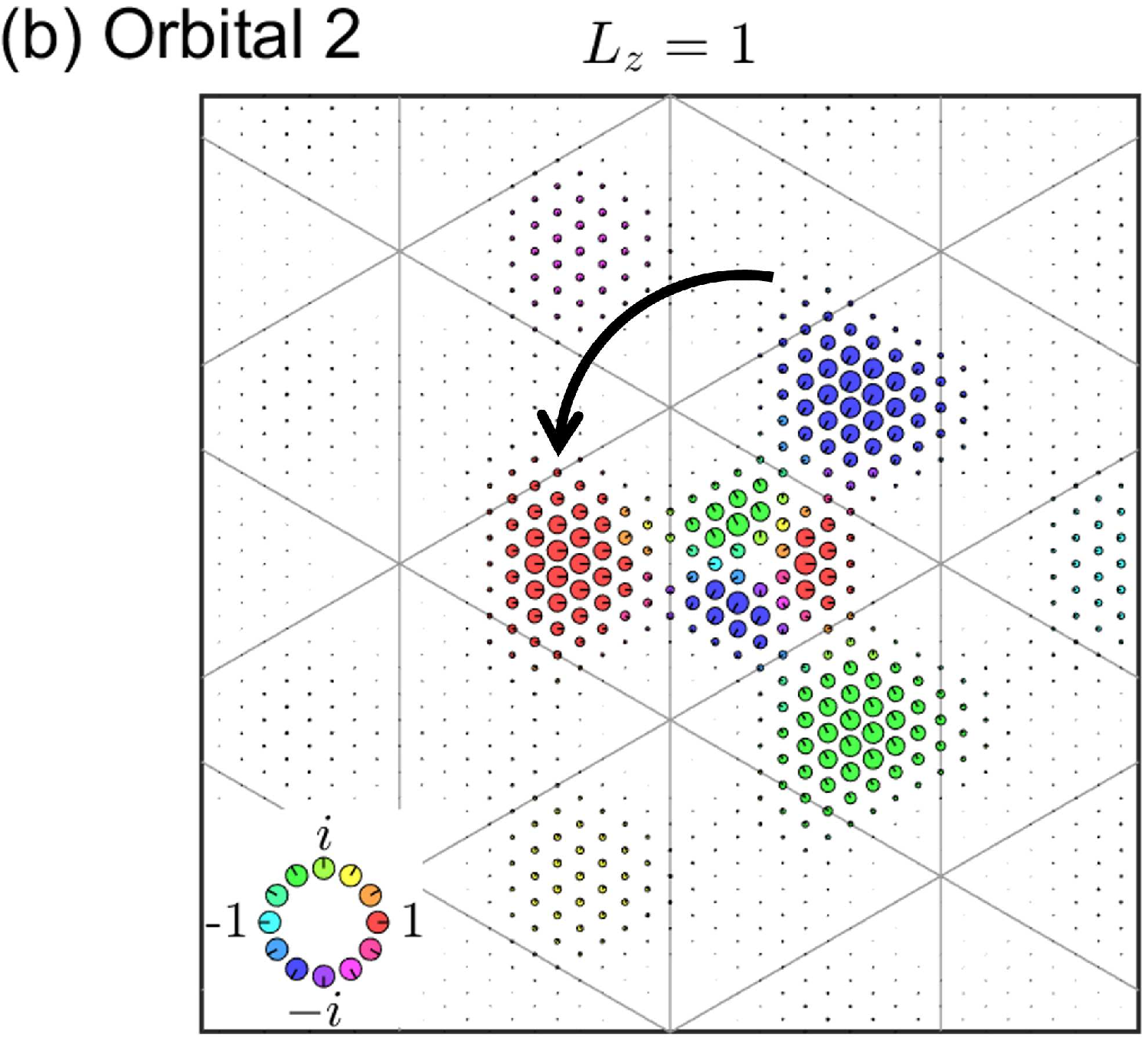}
      \includegraphics[width=0.24\textwidth,valign=t]{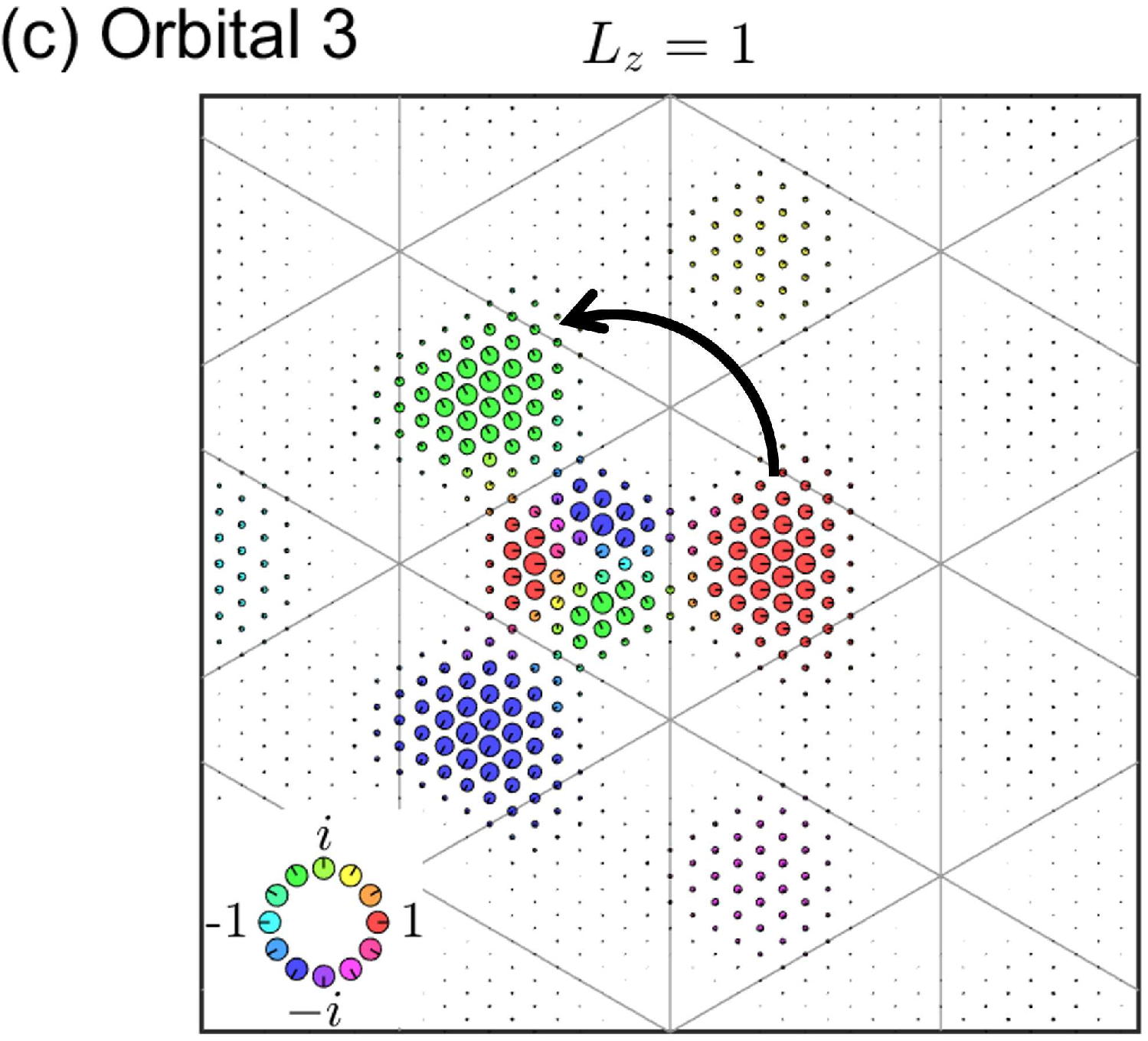}
      \includegraphics[width=0.24\textwidth,valign=t]{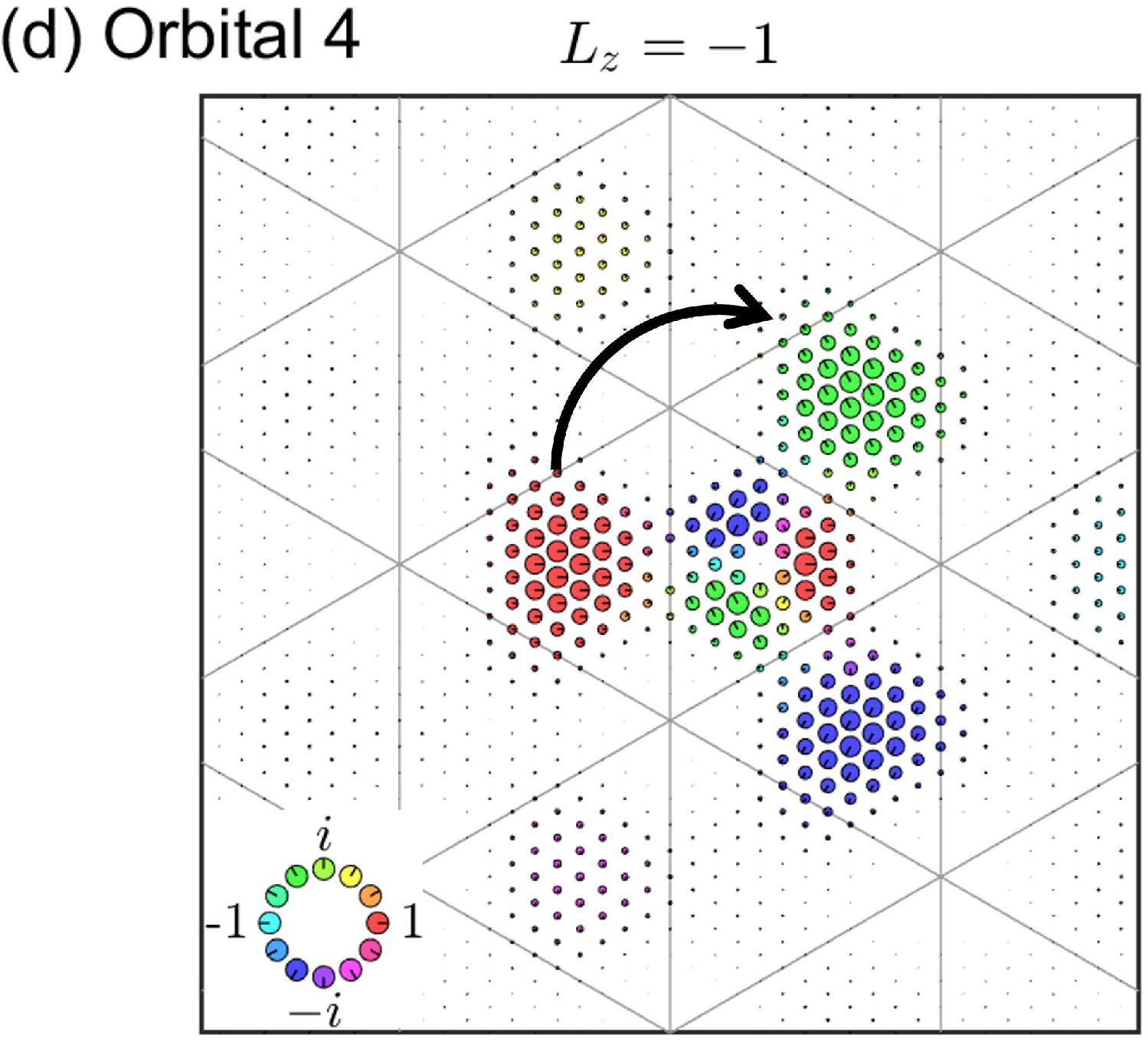} 
\medskip
      \includegraphics[width=0.21\textwidth,valign=t]{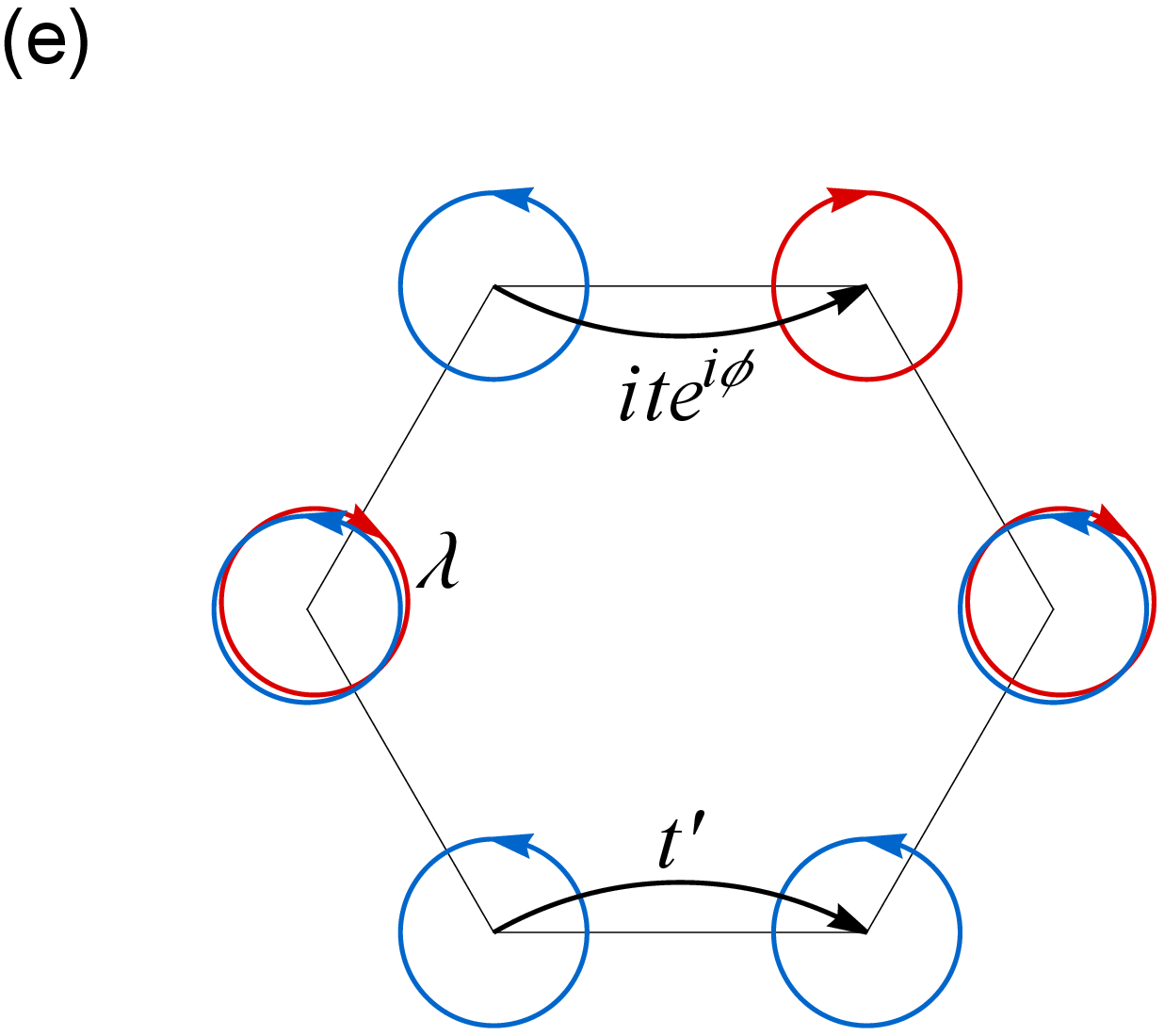}
      \includegraphics[width=0.25\textwidth,valign=t]{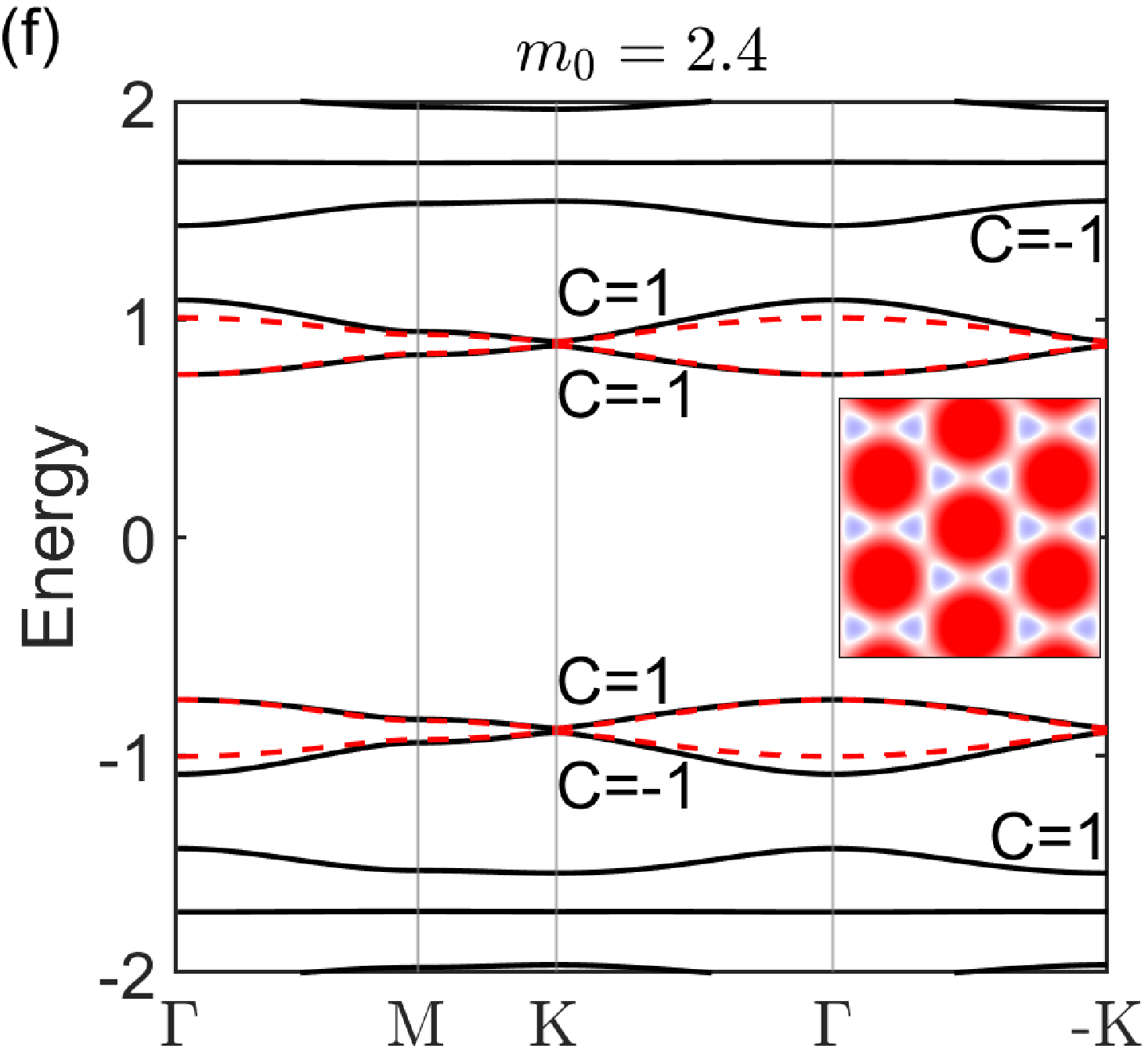} 
      \includegraphics[width=0.25\textwidth,valign=t]{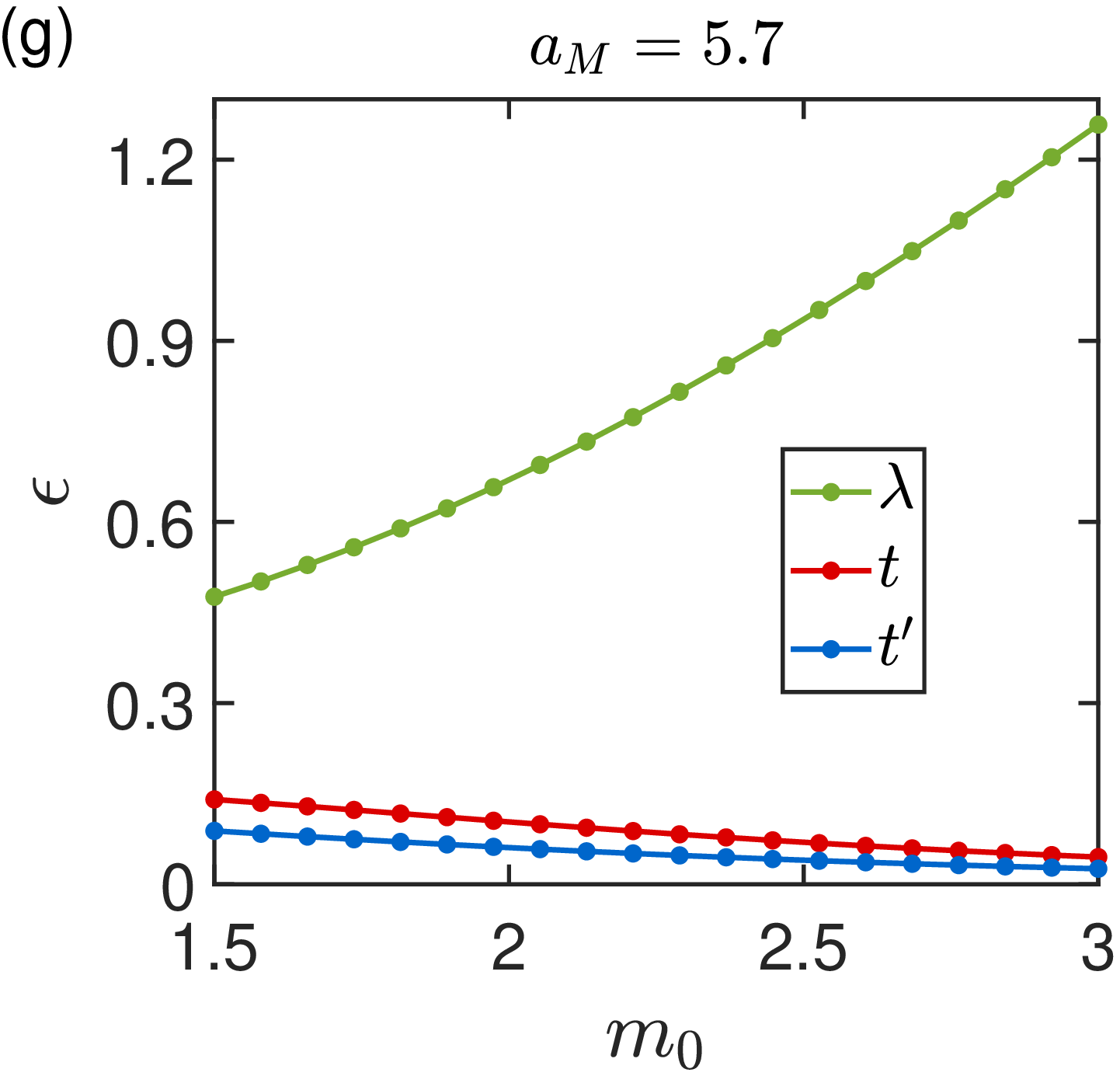} 
      \includegraphics[width=0.25\textwidth,valign=t]{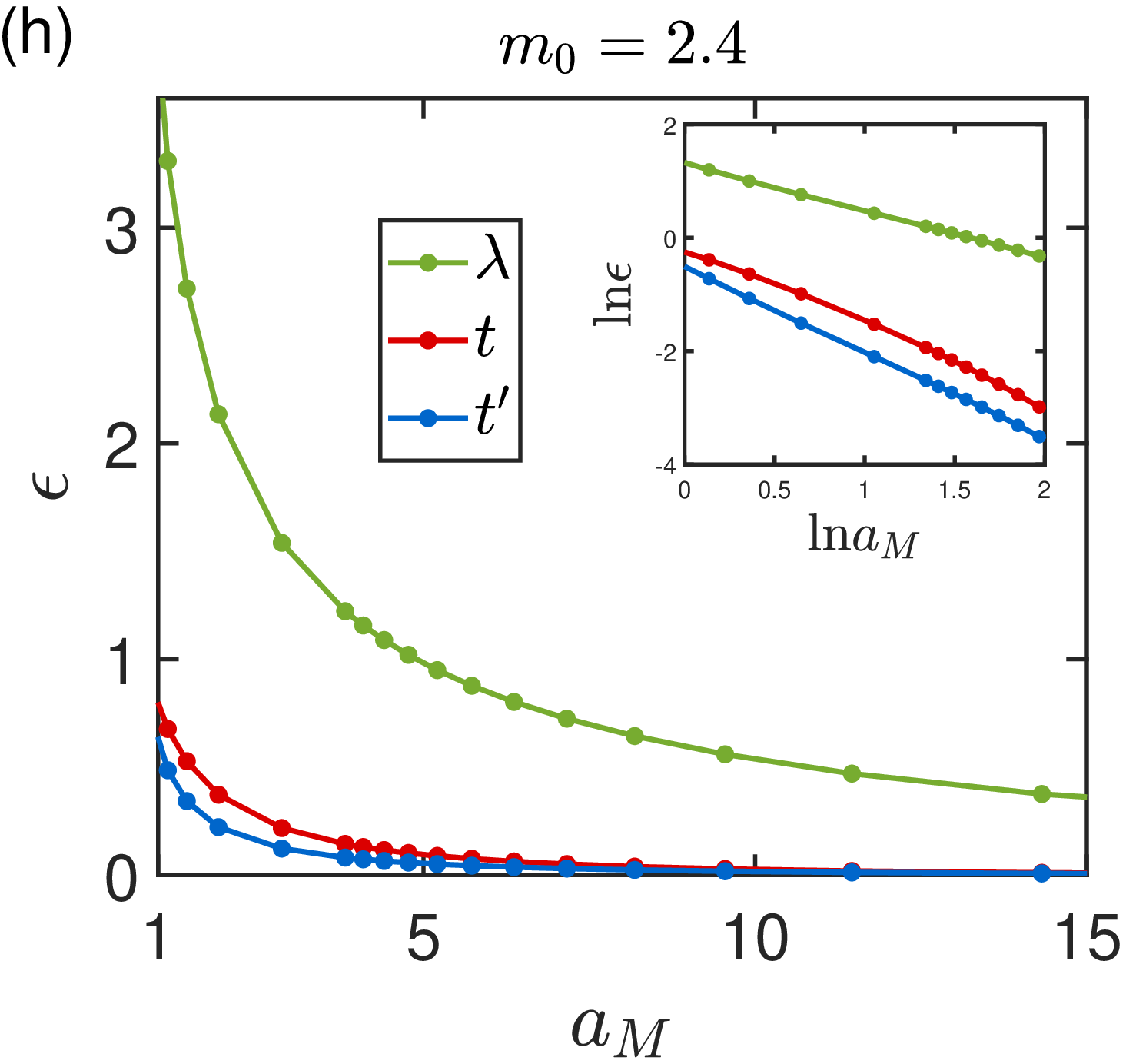} 
    \caption{(a)-(d) Wannier functions for the lowest four lowest bands for $m_0=2.4,a_M=5.7$. Only one component of each Wannier spinor with non-vanishing orbital angular momentum is shown. (e) Sketch of the tight-binding model in Eq. \eqref{eq5} on the honeycomb lattice. (f) Moiré minibands of $H^\uparrow_{\mathbf k}(\mathbf r)$ for $a_M=5.7$ and $m_0=2.4$. The red dashed curves are band structures from the $p_x\pm ip_y$-orbital model on the honeycomb lattice. (g) Hopping amplitudes as a function of $m_0$ for $a_M=5.7$. (h) Hopping amplitudes as a function of $a_M$ for $m_0=2.4$. log-log plot of hopping amplitudes, which have a power-law tail.}
\label{fig7} 
\end{figure*}

\section{Construction of maximally localized Wannier functions}
In this section, we discuss construction of maximally localized Wannier functions. The first step is to find out the localized center of Wannier function and its symmetry. To do this, we plot distributions of Bloch functions at $\mathbf{k}=0$ point for low energy bands as shown in Fig. \ref{fig6}. Each Bloch spinor for the Hamiltonian in Eq. \eqref{eq3} has two components. We define $L_z^{tot}=L_z^1+L_z^2$ as the angular momentum for each band, where $L_z^{i}$ denotes the angular momentum for the $i$-th component of the Bloch spinor. For $m_0=-0.9$, the wave functions are localized on the triangular lattice, so we choose the triangular lattice as the rotation center. Then each Bloch spinor is found to be an eigenstate of $C_{6z}$ with eigenvalue of $e^{i\frac{\pi}{3} L_z^i}$. Hence $(L_z^1,L_z^2)$=(0,1) for the first top occupied band, and $(L_z^1,L_z^2)$=(1,2) for the second top occupied band. Due to the particle-hole symmetry, $(L_z^1,L_z^2)$=(-1,0) for the first bottom unoccupied band, thus $L_z^{tot}=\pm1$ for the lowest two bands. For $m_0=2.4$, the wave functions are localized on the honeycomb lattice, so we choose the honeycomb lattice as the rotation center. Then each Bloch spinor is found to be an eigenstate of $C_{3z}$ with eigenvalue of $e^{i\frac{2\pi}{3} L_z^i}$. Because the lowest two bands as well as the second and third bands have been inverted, their angular momentum are also switched. Then $(L_z^1,L_z^2)$=(-1,0) for the first top occupied band, and $(L_z^1,L_z^2)$=(0,-1) for the second top occupied band. Hence $L_z^{tot}=-1$ for the top two bands and $L_z^{tot}=1$ for the bottom two bands.

Next we prepare the initial wave functions. We note that we separately construct the Wannier representations for varying $m_0$ that changes the localizing sites of electron states. Since the topological phase transition at the CNP involves only the lowest two bands, which are energetically isolated from other bands, we expect two independent Wannier orbitals on the triangular lattice to be transformed from the Bloch functions for these bands. We fix the global phase factor of the Bloch states by different gauges. For the $n$-th Wannier orbital, we fix the phase $e^{i\phi_{n\mathbf{k}}}$ so that $\psi^i_{n\mathbf{k}}(\mathbf{r=0})$ is real and positive, where $i$ denotes the $i$-th component of Bloch spinors with $L_z^{tot}=0$. Then the initial Wannier functions are constructed as
\begin{equation}\label{eq7} 
\begin{aligned}     
    w_{n}(\mathbf r)=\frac{1}{\sqrt{2N}}\sum_{\mathbf{k}}e^{i\phi_{n\mathbf{k}}}\sum_{m=1,2} \psi_{m\mathbf{k}}(\mathbf{r}),
	\end{aligned}
\end{equation}
where the summation in $\mathbf{k}$ is taken over $N$ discrete points in the Brillouin zone. Moreover, we impose the $C_{6z}$ rotational symmetry for the initial Wannier functions. Then we use the maximally localizing method \cite{ref58} to reduce the total spread functional. Here we take $N=18\times18$ in the minimizing procedure. For large $m_0$ we consider the lowest four bands for the Wannierization and take a similar strategy for initial Wannier functions as mentioned above. We note that the chosen gauges allow the initial Wannier trials to be 
well-localized and preserve the desired symmetry as well, so we just take several iterative steps instead of fully minimizing the total spread which 
may break the symmetry of Wannier orbitals.

\section{$p_x\pm ip_y$-orbital model on the honeycomb lattice}
As $m_0>1.38,a_M=5.7$, the second and third miniband are inverted. Using the same approach we construct localized Wannier functions for the lowest four minibands. As shown in Fig. \ref{fig7}(a)-(d), the four Wannier functions characterize the chiral $p_x\pm ip_y$ orbitals on the honeycomb lattice. The tight-binding model in Wannier representations has the same formula with Eq. \eqref{eq5},
where the on-site orbital rotation now plays a role as an effective SOC \cite{ref30}. The energy dispersion and band topology are both in good agreement with the band structures from the continuum model, as shown in Fig. \ref{fig7}(f). Since the honeycomb lattice is right at the center of each triangular plaquette, the above mentioned neighboring hopping on the triangular lattice turns into the effective SOC on the honeycomb lattice, and vice versa. As a result, as $m_0$ increases, $\lambda$ increases while $t$ and $t'$ decrease, as shown in Fig. \ref{fig7}(g), which is inverse of Fig. \ref{fig3}(e). The effective SOC is larger than the neighboring hopping, and the CNP gap between the first and second band is inverse of $\lambda$ or $m_0$ \cite{ref31,ref52}. Similarly, the hopping amplitudes have a power-law decay respect to $a_M$, as shown in Fig. \ref{fig7}(f).

\bibliographystyle{aipnum4-1}
\bibliography{ref}

\end{document}